\documentclass[10pt, journal,letterpaper]{IEEEtran}
\pdfoutput=1
\usepackage{geometry}
\usepackage{hyperref}
\hypersetup{
	pdfinfo={
		Title={QECO: A QoE-Oriented Computation Offloading Algorithm based on Deep Reinforcement Learning for Mobile Edge Computing},
		Author={Iman Rahmaty, Hamed Shah-Mansouri, Ali Movaghar},
		Subject={IEEE Internet of Things Journal},
		Keywords={IoT, quality of experience, deep reinforcement learning, computation task offloading, Mobile edge computing}
	}
}

\usepackage{pbalance}

\usepackage{amsmath}

\usepackage{upquote}

\usepackage[ngerman,main=english]{babel}
\addto\extrasenglish{\languageshorthands{ngerman}\useshorthands{"}}

\makeatletter
\g@addto@macro{\UrlBreaks}{\UrlOrds}
\makeatother

\usepackage[zerostyle=b,scaled=.75]{newtxtt}

\usepackage[T1]{fontenc}

\usepackage[
  babel=true, 
  expansion=alltext,
  protrusion=alltext-nott, 
  final 
]{microtype}

\DisableLigatures{encoding = T1, family = tt* }

\usepackage{graphicx}

\usepackage{diagbox}

\usepackage{xcolor}

\usepackage{listings}

\definecolor{eclipseStrings}{RGB}{42,0.0,255}
\definecolor{eclipseKeywords}{RGB}{127,0,85}
\colorlet{numb}{magenta!60!black}

\usepackage{multirow}

\renewcommand{\arraystretch}{1.2}

\lstdefinelanguage{json}{
    basicstyle=\normalfont\ttfamily,
    commentstyle=\color{eclipseStrings}, 
    stringstyle=\color{eclipseKeywords}, 
    numbers=left,
    numberstyle=\scriptsize,
    stepnumber=1,
    numbersep=8pt,
    showstringspaces=false,
    breaklines=true,
    frame=lines,
    string=[s]{"}{"},
    comment=[l]{:\ "},
    morecomment=[l]{:"},
    literate=
        *{0}{{{\color{numb}0}}}{1}
         {1}{{{\color{numb}1}}}{1}
         {2}{{{\color{numb}2}}}{1}
         {3}{{{\color{numb}3}}}{1}
         {4}{{{\color{numb}4}}}{1}
         {5}{{{\color{numb}5}}}{1}
         {6}{{{\color{numb}6}}}{1}
         {7}{{{\color{numb}7}}}{1}
         {8}{{{\color{numb}8}}}{1}
         {9}{{{\color{numb}9}}}{1}
}

\usepackage[autostyle=true]{csquotes}

\defineshorthand{"`}{\openautoquote}
\defineshorthand{"'}{\closeautoquote}

\usepackage{booktabs}

\usepackage{paralist}

\usepackage[%
  square,        
  comma,         
  numbers,       
  sort&compress 
]{natbib}

\usepackage{etoolbox}
\makeatletter
\patchcmd{\NAT@test}{\else \NAT@nm}{\else \NAT@hyper@{\NAT@nm}}{}{}
\makeatother

\usepackage{pdfcomment}

\usepackage{stfloats}
\fnbelowfloat

\usepackage[group-minimum-digits=4,per-mode=fraction]{siunitx}
\addto\extrasgerman{\sisetup{locale = DE}}

\usepackage{hyperref}

\usepackage{afterpage}

\usepackage[all]{hypcap}

\usepackage[caption=false,font=footnotesize]{subfig}

\usepackage[incolumn]{mindflow}

\usepackage[capitalise,nameinlink,noabbrev]{cleveref}

\crefname{listing}{Listing}{Listings}
\Crefname{listing}{Listing}{Listings}
\crefname{lstlisting}{Listing}{Listings}
\Crefname{lstlisting}{Listing}{Listings}

\usepackage{lipsum}

\usepackage[math]{blindtext}
\usepackage{mwe}
\usepackage[realmainfile]{currfile}
\usepackage{tcolorbox}

\usepackage{Carrickc,lettrine}

\tcbuselibrary{listings}

\DeclareFontFamily{U}{MnSymbolC}{}
\DeclareSymbolFont{MnSyC}{U}{MnSymbolC}{m}{n}
\DeclareFontShape{U}{MnSymbolC}{m}{n}{
  <-6>    MnSymbolC5
  <6-7>   MnSymbolC6
  <7-8>   MnSymbolC7
  <8-9>   MnSymbolC8
  <9-10>  MnSymbolC9
  <10-12> MnSymbolC10
  <12->   MnSymbolC12%
}{}
\DeclareMathSymbol{\powerset}{\mathord}{MnSyC}{180}

\usepackage{xspace}

\makeatletter
\newcommand{\hydash}{\penalty\@M-\hskip\z@skip}
\makeatother

\input glyphtounicode
\usepackage{bbm}
\usepackage{amssymb}

\usepackage{algorithm}
\usepackage{algpseudocode}

\usepackage{amsmath}

\usepackage{tikz,xcolor,hyperref}

\definecolor{lime}{HTML}{A6CE39}
\DeclareRobustCommand{\orcidicon}{%
	\begin{tikzpicture}
		\draw[lime, fill=lime] (0,0) 
		circle [radius=0.14] 
		node[white] {{\fontfamily{qag}\selectfont \tiny ID}};
		\draw[white, fill=white] (-0.0625,0.095) 
		circle [radius=0.007];
	\end{tikzpicture}
	\hspace{-3mm}
}

\foreach \x in {A, ..., Z}{%
	\expandafter\xdef\csname orcid\x\endcsname{\noexpand\href{https://orcid.org/\csname orcidauthor\x\endcsname}{\noexpand\orcidicon}}
}

\usepackage{blindtext,graphicx}
\usepackage[absolute]{textpos}
\usepackage{hyperref}
\hypersetup{
	colorlinks = true,
	anchorcolor = {blue},
	citecolor = {blue},
	linkcolor = {blue},	
	urlcolor={blue}
}

\newcommand{\Cite}[1]{\textcolor{blue}{\cite{#1}}}

\makeatletter
\renewcommand\@cite[2]{\textcolor{blue}{[{#1\if@tempswa , #2\fi}]}}
\makeatother

\usepackage{amssymb}
\usepackage{pifont}
\newcommand{\cmark}{\ding{51}}%
\newcommand{\xmark}{\ding{55}}%

\begin{document}



\title{QECO: A QoE-Oriented Computation Offloading Algorithm based on Deep Reinforcement \\Learning for Mobile Edge Computing}

\author{\IEEEauthorblockN{
		Iman Rahmaty\orcidA{}, Hamed Shah-Mansouri\orcidB{}, \textit{Member, IEEE}, and Ali Movaghar\orcidC{}, \textit{Life Senior Member, IEEE}}  

\thanks{Received 27 March 2024; revised 28 November 2024; accepted 25 March 2025. Date of publication 1 April 2025; date of current version 27 June 2025. Recommended for acceptance by Dr. Hongbo Jiang. \textit{(Corresponding author: Hamed Shah-Mansouri.)}}
\thanks{Iman Rahmaty and Ali Movaghar are with the Department of Computer Engineering, Sharif University of Technology, Tehran 14588-89694, Iran (e-mail: \href{mailto:iman.rahmati@sharif.edu}{iman.rahmati@sharif.edu}; \href{mailto:movaghar@sharif.edu}{movaghar@sharif.edu}).}
\thanks{Hamed Shah-Mansouri is with the Department of Electrical Engineering, Sharif University of Technology, Tehran 14588-89694, Iran (e-mail: \href{mailto:hamedsh@sharif.edu}{hamedsh@sharif.edu}).}
\thanks{Digital Object Identifier \href{https://ieeexplore.ieee.org/document/10946841}{10.1109/TNSE.2025.3556809}
}
\thanks{Source code available at \href{https://github.com/ImanRHT/QECO}{https://github.com/ImanRHT/QECO}
}
}


\maketitle

%
%

\begin{abstract}
In the realm of mobile edge computing (MEC), efficient computation task offloading plays a pivotal role in ensuring a seamless quality of experience (QoE) for users. Maintaining a high QoE is paramount in today's interconnected world, where users demand reliable services. This challenge stands as one of the most primary key factors contributing to handling dynamic and uncertain mobile environments. In this study, we delve into computation offloading in MEC systems, where strict task processing deadlines and energy constraints can adversely affect the system performance. We formulate the computation task offloading problem as a Markov decision process (MDP) to maximize the long-term QoE of each user individually. We propose a distributed QoE-oriented computation offloading (QECO) algorithm based on deep reinforcement learning (DRL) that empowers mobile devices to make their offloading decisions without requiring knowledge of decisions made by other devices. Through numerical studies, we evaluate the performance of QECO.
	Simulation results reveal that compared to the state-of-the-art existing works, QECO increases the number of completed tasks by up to 14.4\%, while simultaneously reducing task delay and energy consumption by 9.2\% and 6.3\%, respectively. Together, these improvements result in a significant average QoE enhancement of 37.1\%. This substantial improvement is achieved by accurately accounting for user dynamics and edge server workloads when making intelligent offloading decisions. This highlights QECO's effectiveness in enhancing users' experience in MEC systems.


\end{abstract}


%
\IEEEpeerreviewmaketitle


%
%

\thispagestyle{empty}
\begin{textblock}{19.1}(1,1.4)
	\noindent \scriptsize \hspace{4mm}
	3118 \hfill IEEE TRANSACTIONS ON NETWORK SCIENCE AND ENGINEERING, VOL. 12, NO. 4, JULY/AUGUST 2025
\end{textblock}

\begin{textblock}{19.1}(1,1.5)
	\vspace{25.4cm}
	\noindent \footnotesize \hspace{10mm}
	2327-4697 © 2025 IEEE. All rights reserved, including rights for text and data mining, and training of artificial intelligence and similar technologies.\\ .\hspace{5mm} Personal use is permitted, but republication/redistribution requires IEEE permission. See https://www.ieee.org/publications/rights/index.html for more information.	
\end{textblock}

\vspace{-2mm}

\section{Introduction} 

\lettrine{\fontsize{33}{33}\textbf{M}}\,\,OBILE edge computing (MEC)~\Cite{mao2017survey} has emerged as~a promising technological solution to overcome the challenges faced by mobile devices (MDs) when performing high computational tasks, such as real-time data processing and artificial intelligence applications~\Cite{zhou2019edge}. In spite of the MDs' technological advancements, their limited computing power and energy may lead to task drops, processing delays, and an overall poor user experience. By offloading intensive tasks to nearby edge nodes (ENs), MEC effectively empowers computation capability and reduces the delay and energy consumption. This improvement enhances the users' quality of experience (QoE), especially for time-sensitive computation tasks~\Cite{TNSE-QOE-24},~\Cite{shah2018hierarchical}. 

Efficient task offloading in MEC is a complex optimization challenge due to the dynamic nature of the network and the variety of MDs and servers involved~\Cite{jiang2019toward},~\Cite{TNSE-WU-24}. In particular, determining the optimal offloading strategy, scheduling the tasks, and selecting the most suitable EN for task offloading are the main challenges that demand careful consideration. In addition to the dynamic network changes, the uncertain requirements and sensitive latency properties of computation tasks pose nontrivial challenges that can significantly impact the MEC systems.

To cope with the dynamic nature of the network, recent research has proposed several task offloading algorithms using machine learning methods. In particular, reinforcement learning (RL) \Cite{mnih2015human} holds promises to determine optimal decision-making policies by capturing the dynamics of environments and learning strategies for accomplishing long-term objectives. However, the traditional RL methods are not efficient in handling environments with high-dimensional state  spaces. Deep reinforcement learning (DRL) combines  traditional RL  with  deep  neural  networks  to  intelligently respond to the unknown and dynamic system when addressing the limitations of RL-based algorithms. Despite these advancements, task offloading still faces significant challenges in real-world scenarios with multiple MDs and ENs. In such scenarios, it is essential for MDs to make offloading decisions independently  without prior knowledge of other MDs’ tasks and offloading models. However, existing works fail to adequately address this challenge. In addition, QoE is a time-varying performance measure that reflects user satisfaction and is not affected only by delay, but also by energy consumption. Albeit some existing works have investigated the trade-off between delay and energy consumption, they fail to properly estimate the QoE value and address the users' requirements. Although DRL offers a valuable tool, a more comprehensive approach is required to accurately estimate the QoE while addressing the aforementioned challenges in practical scenarios where global information is limited.

In this study, we delve into the computation task offloading problem in MEC systems, where strict task processing deadlines and energy constraints can adversely affect the system performance. We propose a distributed QoE-oriented computation offloading (QECO) algorithm to efficiently handle task offloading under uncertain loads at ENs. This algorithm empowers MDs to make offloading decisions utilizing only locally observed information, such as task size, queue details, battery status, and historical workloads at the ENs. To capture the dynamic nature of the MEC environment, we employ the dueling double deep Q-network (D3QN), which is a refined improvement over standard deep Q-network (DQN) model. By integrating both double Q-learning \Cite{van2016deep} and dueling network architectures \Cite{wang2016dueling}, D3QN reduces overestimation bias in action-value predictions and more accurately distinguishes the relative importance of states and actions. We also incorporate long short-term memory (LSTM) \Cite{hochreiter1997long} into our D3QN model to continuously estimate dynamic workloads at ENs. This enables MDs to effectively handle the uncertainty of the MEC environment, where global information is limited, and proactively adjust the offloading strategies to improve their long-term QoE estimation. By adopting the appropriate policy based on each MD’s specific requirements, the~QECO algorithm significantly improves the QoE for individual users. 

Our main contributions are summarized as follows:
\begin{itemize}
	\item \textit{Task Offloading Problem in the MEC System:} We consider queuing systems for MDs and ENs and formulate the task offloading problem while accounting for the time-varying system environments (e.g., the arrival of new tasks, and the computational requirement of each task). We aim to maximize the long-term QoE, which reflects the task completion rate, task delay, and energy consumption of MDs. Our problem formulation effectively utilizes the resources and properly handles the dynamic fluctuations of workload at ENs.

	 \item \textit{DRL-based Offloading Algorithm:} To solve the problem of long-term QoE maximization in highly dynamic mobile environments, we propose the QECO algorithm based on DRL, which empowers each MD to make offloading decisions independently, without prior knowledge of other MDs' tasks and offloading models. Focusing on the MD's QoE preference, our approach leverages D3QN and LSTM to prioritize and strike an appropriate balance between QoE factors while accounting for workload uncertainty at the ENs. QECO empowers MDs to anticipate the EN's load level over time, leading to more accurate offloading strategies. We also analyze the training convergence and computational complexity of the proposed algorithm.

	\item \textit{Performance Evaluation:} We conduct comprehensive experiments to evaluate QECO’s performance and its training convergence against baseline schemes as well as two state-of-the-art existing works. The results demonstrate that our algorithm effectively utilizes the computing resources while taking the dynamic workloads at ENs into consideration. In particular, QECO converges more quickly than vanilla DQN and double DQN (DDQN) methods. QECO also improves average QoE by 37.1\% and 29.8\% compared to the potential game-based offloading algorithm (PGOA)~\Cite{yang2018distributed} and distributed and collective DRL-based offloading algorithm (DCDRL)~\Cite{qiu2020distributed}, respectively.
\end{itemize}

 The structure of this paper is as follows. Section~\ref{section:II} reviews the related works. Section~\ref{section:III} presents the system model, followed by the problem formulation in Section ~\ref{section:IV}. In Section~\ref{section:V}, we present the algorithm, while Section~\ref{section:VI} provides an evaluation of its performance. Finally, we conclude in Section~\ref{section:VII}. \color{black}

\thispagestyle{empty}
\begin{textblock}{19.1}(1,1.4)
	\vspace{-5mm}
	\noindent \scriptsize \hspace{4mm}
	RAHMATY et al.: QECO: A QOE-ORIENTED COMPUTATION OFFLOADING ALGORITHM BASED ON DEEP REINFORCEMENT LEARNING \hfill 3119
\end{textblock}

\color{black}
\section{Related Work}
\label{section:II}

To effectively tackle the challenges of MEC arising from the ever-changing nature of networks, recent research highlights the effectiveness of RL in adapting to environmental changes and learning optimal strategies. In this section, we explore the current RL-based state-of-the-art works and discuss their strengths and limitations. Table~\ref{table1} provides an intuitive comparison of these works.
\begin{table*}[tbp]\centering
		\renewcommand{\arraystretch}{1}
		\captionsetup{name=TABLE}
		\caption{Comparison of Related Works}
		\scalebox{1}{
			\begin{tabular}{ lp{4.4cm}p{0.5cm}p{0.6cm}p{0.4cm}p{0.6cm}p{2.3cm}p{3.7cm}l} 
				\toprule
				\multirow{2}{*}{{\textbf{Paper}}}&\multirow{2}{*}{{\textbf{Problem}}} &\multicolumn{4}{@{}l@{}}{\textbf{Optimization target}}&\multirow{2}{*}{{\textbf{Method}}} & \multirow{2}{*}{{\textbf{Limitation}}}   \\ \cmidrule{3-6}
				&&Delay&Energy& Drop& QoE &&&\vspace{1mm}  \\ 							\toprule
				\Cite{zhang2023offline}  & Computation offloading &  \centering \cmark & \centering \xmark&  \centering \xmark&  \centering \xmark& Offline RL & \multirow{6}{*}{\shortstack[l]{Overlooking energy consumption; \\Not suitable for energy-constrained\\ MEC; May increase energy \\consumption or reduce the overall\\ system capacity.}} \\ \cmidrule{1-7}
				\Cite{sun2024hierarchical}  & Joint offloading and service caching & \centering \cmark & \centering \xmark&  \centering \xmark&  \centering \xmark& Hierarchical DRL &   \\ \cmidrule{1-7}
				\Cite{liu2022deep} & Computation offloading  &  \centering \cmark &  \centering \xmark &  \centering \xmark&  \centering \xmark& DQN, DDPG&    \\  \cmidrule{1-7}
				\Cite{tang2022double} & Computation offloading &  \centering \cmark & \centering \xmark&  \centering \xmark&  \centering \xmark& DDQN &   \\  \cmidrule{1-7}
				\Cite{li2022integrated} &  Joint offloading and resource allocation &  \centering \cmark & \centering \xmark&  \centering \xmark&  \centering \xmark& AC network&   \\  \cmidrule{1-7}
				\Cite{wei2023many} & Computation offloading  &  \centering \cmark & \centering \xmark&  \centering \cmark&\centering\xmark & AC network  &&\vspace{1mm}  \\  								\toprule
				\Cite{munir2021multi}  & Computation offloading & \centering\xmark &\centering\cmark&\centering \xmark& \centering\xmark& Multi-agent RL & \multirow{6}{*}{\shortstack[l]{Do not account for time-sensetive\\ applications; Reducing energy\\ consumption can increase \\computation delay, potentially \\resulting in task failures.}}  \\ \cmidrule{1-7}
				\Cite{zhou2021deep}  & Joint offloading and resource allocation & \centering\xmark &\centering\cmark& \centering\xmark& \centering\xmark& DDQN &   \\ \cmidrule{1-7}
				\Cite{dai2020edge} & Joint offloading and resource allocation & \centering\xmark &\centering\cmark& \centering\xmark&\centering\xmark& DDPG &   \\  \cmidrule{1-7}
				\Cite{chouikhi2023energy} & Computation offloading & \centering\xmark &\centering\cmark&\centering \cmark&\centering\xmark& DRL &   \\  \cmidrule{1-7}
				\Cite{wu2024combining} & Privacy aware computation offloading &\centering \xmark &\centering\cmark&\centering \cmark&\centering\xmark&AC network && \vspace{1mm}  \\  	\toprule
				\Cite{huang2021deadline} & Computation offloading  & \centering\cmark &\centering\cmark& \centering\xmark&\centering\xmark& DDPG&\multirow{6}{*}{\shortstack[l]{Not suitable for applications with \\strict delay requirements. Tasks may \\be dropped if there is a strict \\requirement on either objective.}} \\  			 \cmidrule{1-7}
				\Cite{liao2023online} & Online computation offloading & \centering\cmark &\centering\cmark& \centering\xmark& \centering\xmark&DDQN& \\  			 \cmidrule{1-7}
				\Cite{wu2023multi} & Computation offloading & \centering\cmark &\centering\cmark& \centering\xmark&\centering\xmark& Multi-agent PPO&  \\  			 \cmidrule{1-7}
				\Cite{gong2022edge} & Joint offloading and resource allocation & \centering\cmark &\centering\cmark& \centering\xmark&\centering\xmark& DRL &   \\  			 \cmidrule{1-7}
				\Cite{liu2021learn}  &  Joint offloading and resource allocation  & \centering\cmark &\centering\cmark& \centering\xmark&\centering\xmark & Multi-agent AC &   \\  \cmidrule{1-7}
				\Cite{she2024efficient} & Computation offloading & \centering\cmark &\centering\cmark& \centering\xmark&\centering\xmark& Multi-agent DDPG &&\vspace{1mm}    \\  						\toprule
				\Cite{gao2022large} & Computation offloading & \centering\cmark &\centering\cmark& \centering\cmark& \centering\xmark& Multi-agent AC&\multirow{6}{*}{\shortstack[l]{Overlooking personalized QoE; \\Treating all MDs similarly when \\optimzing the overall system \\performance fails to account for \\individual user satisfaction.}}  \\  			 \cmidrule{1-7}
				\Cite{wu2024privacy} & Privacy preserving offloading & \centering\cmark &\centering\cmark& \centering\cmark&\centering\xmark& Multi-agent AC  &   \\  			 \cmidrule{1-7}
				\Cite{Bolourian-WCL24} & Energy harvesting offloading & \centering\cmark &\centering\cmark& \centering\cmark&\centering\xmark& DQN&   \\  			 \cmidrule{1-7}
				\Cite{wu2023computation} &Joint offloading and resource allocation  & \centering\cmark &\centering\cmark&\centering\cmark& \centering\xmark& Multi-agent RL & &\vspace{1mm}  \\  										\toprule
				QECO & QoE-oriented computation offloading  & \centering\cmark &\centering\cmark& \centering\cmark&\centering\cmark& D3QN + LSTM &&\vspace{1mm} 	 		  \\
				\toprule
		\end{tabular}}
		\label{table1}
\end{table*}

For time-sensitive applications, development of delay-optimal approaches is the primary goal. 
To solve the delay minimization problem in MEC environment, Zhang \textit{et al.} in \Cite{zhang2023offline} proposed an offline RL-based computation offloading algorithm, where the problem is modeled as a repeated game between two agents.
Sun \textit{et al.} in \Cite{sun2024hierarchical} explored both computation offloading and service caching problems in MEC. They formulated an optimization problem that aims to minimize the long-term average service delay. They then proposed a hierarchical DRL framework, which effectively handles both problems under heterogeneous resources. Liu \textit{et al.} in \Cite{liu2022deep} formulated the offloading problem as an MDP and proposed a deep deterministic policy gradient (DDPG)-based approach to minimize both communication and computation delay in MEC.
To minimize total delay and reduce mobile vehicle task waiting time, Tang \textit{et al.} in \Cite{tang2022double} developed a dynamic offloading model for multiple vehicles, segmenting tasks into sequential subtasks for more precise offloading decisions.
Li \textit{et al.} in \Cite{li2022integrated} proposed an actor-critic (AC)-based algorithm to optimize computation offloading and resource allocation decisions, which leverages past experience and model knowledge to enable fast and resilient real-time offloading control. 
In \Cite{wei2023many}, the authors investigated deadline-constrained time-sensitive tasks in vehicular MEC networks and proposed a multi-agent RL-based computation offloading algorithm to address joint delay and computation rate optimization. Despite their benefits, the delay-optimal approaches~\Cite{zhang2023offline}--\Cite{wei2023many} often sacrifice other key performance metrics and may increase energy consumption or diminish overall computational capacity.

For energy-constrained MEC systems, reducing energy consumption becomes the key objective, motivating the development of energy-efficient approaches. 
Munir \textit{et al.} in \Cite{munir2021multi} investigated an energy dispatch problem for a self-powered MEC system and proposed a semi-distributed approach using a multi-agent RL framework to minimize energy consumption.
Zhou \textit{et al.} in \Cite{zhou2021deep} proposed a Q-learning approach, which is an extension of RL to achieve optimal resource allocation strategies and computation offloading.
Dai \textit{et al.} in \Cite{dai2020edge} introduced the integration of action refinement into DRL and designed an algorithm based on DDPG to optimize resource allocation and computation offloading concurrently. 
To ensure that minimizing energy consumption does not lead to a reduction in the system's computing rate, researchers have expanded their work by incorporating maximum tolerable delay constraints.
In \Cite{chouikhi2023energy}, Chouikhi \textit{et al.} proposed a DRL-based computation offloading approach that aims to minimize long-term energy consumption and maximize the number of tasks completed within their deadlines.
To address the privacy-aware computation offloading problem, Wu \textit{et al.} in \Cite{wu2024combining} proposed a DQN-based method to optimize the computation rate and energy consumption in a queuing-based Industrial Internet of Things (IIoT) network. 
Huang \textit{et al.} in \Cite{huang2021deadline} proposed a DRL-based method using partially observable MDP (POMDP), which ensures the real-time tasks deadlines are met while minimizing the total energy consumption of MDs. This algorithm effectively tackles the challenges of dynamic resource allocation in large-scale heterogeneous networks. 
The aforementioned energy-efficient approaches~\Cite{munir2021multi}--\Cite{huang2021deadline} do not adequately address the requirements of time-sensitive applications, where reducing delay could have significant effects on the users' QoE.

Energy consumption and computation delay are two conflicting objectives that must be properly balanced when dealing with time-sensitive applications in energy-constrained MEC systems.
Liao \textit{et al.} in \Cite{liao2023online} introduced a DDQN-based algorithm for performing online computation offloading in MEC. This algorithm optimizes transmission power and CPU frequency when minimizing both task computation delay and energy consumption. 
To optimize delay and energy consumption, Wu \textit{et al.} in \Cite{wu2023multi} investigated the computation offloading problem in a queuing-based MEC IIoT system. They modeled the problem as a POMDP and proposed a multi-agent proximal policy optimization (PPO)-based method to obtain the optimal offloading strategy in dynamic environments. 
Gong \textit{et al.} in \Cite{gong2022edge} proposed a DRL-based algorithm jointly optimizes task offloading and resource allocation to achieve lower energy consumption and decreased task delay in IIoT systems.
Liu \textit{et al.} in \Cite{liu2021learn} investigated a two-timescale computing offloading and resource allocation problem and proposed a resource coordination algorithm based on multi-agent DRL, which can generate interactive information along with resource allocation decisions. 
She \textit{et al.} in \Cite{she2024efficient} formulated the computation offloading problem as a POMDP and proposed a DRL-based approach to minimize both transmission delay and energy.
Although delay-energy trade-off approaches~\Cite{liao2023online}--\Cite{she2024efficient} effectively address the needs of time-sensitive tasks, they may lead to task failures and drop when strict requirements are imposed on one of the objectives.

\thispagestyle{empty}
\begin{textblock}{19.1}(1,1.4)
	\vspace{-5mm}
	\noindent \scriptsize \hspace{4mm}
	3120 \hfill IEEE TRANSACTIONS ON NETWORK SCIENCE AND ENGINEERING, VOL. 12, NO. 4, JULY/AUGUST 2025
\end{textblock}

In addition to energy and delay, task computation rate is also of paramount importance. An efficient offloading approach should jointly consider these factors and achieve a proper trade-off among them.
In \Cite{gao2022large}, Gao \textit{et al.} introduced an attention-based multi-agent algorithm designed for decentralized computation offloading that jointly considers the aforementioned objectives.
To optimize privacy protection and quality of service, authors in \Cite{wu2024privacy} investigated the joint computation offloading and power allocation problems for an IIoT network. They modeled the problem as an MDP and proposed a multi-agent DQN-based algorithm. 
In \Cite{Bolourian-WCL24}, the authors proposed an offloading algorithm using DQN for wireless-powered IoT devices in MEC. This algorithm aims to minimize the task drop rate while the devices solely rely on harvested energy for operation. 
Wu \textit{et al.} in \Cite{wu2023computation} introduced a stochastic game-based resource allocation in the MEC. They used an MDP and proposed a multi-agent RL method with the goal of minimizing both energy consumption and delay. Studies~\Cite{gao2022large}--\Cite{wu2023computation} have examined delay-sensitive applications with processing deadlines and addressed energy-constrained MEC challenges, but they overlook user-specific requirements and fail to effectively meet QoE expectations.

The aforementioned existing works have advanced computation offloading in MEC systems by targeting various objectives, which are usually conflicting. These approaches typically compromise one or a subset of objectives to meet their targets. However, while promising, they may not be suitable for time-sensitive applications in energy-constrained MEC environments, especially under the stringent requirements of MDs and the highly dynamic loads of ENs. For instance, studies~\Cite{zhang2023offline}--\Cite{huang2021deadline} focus solely on energy consumption or computation delay. While~\Cite{liao2023online}--\Cite{she2024efficient} jointly consider these two targets, they neglect task completion and fail to prevent task failures. Recent works \Cite{gao2022large}--\Cite{wu2023computation} attempt to address these challenges by incorporating energy, delay, and task completion. However, they disregard users' QoE requirements and their individual preferences. QoE is a time-varying performance measure that reflects user satisfaction, with each user prioritizing its specific needs over overall system performance. These gaps motivate us to design an adaptive algorithm to balance conflicting objectives and enhance QoE for MEC users individually.

In addition to the above limitations, the constrained computing recourses of ENs necessitate careful attention. Existing works either neglect the resource constraints (e.g., \Cite{chen2021drl}) or fail to properly account for the ENs' dynamic load, particularly under limited information (e.g.,~\Cite{sun2024hierarchical},~\Cite{dai2020edge},~\Cite{gong2022edge}, \Cite{gao2022large},~\Cite{wu2024privacy},~\Cite{zhao2019deep}). These lead to degraded system performance. Different from these works, we propose a distributed algorithm that effectively manages unknown load dynamics at ENs. Our algorithm enables each MD to make offloading decisions independently, without requiring information (e.g., task models, offloading decisions) from other MDs, thereby reducing signaling overhead and improving overall system performance.

\color{black}

\section{System Model} 
\label{section:III}
\label{sec:latexhints}

 \begin{figure}
	\captionsetup{name=Fig.}
	\centering
	\includegraphics[width=0.9\linewidth]{ 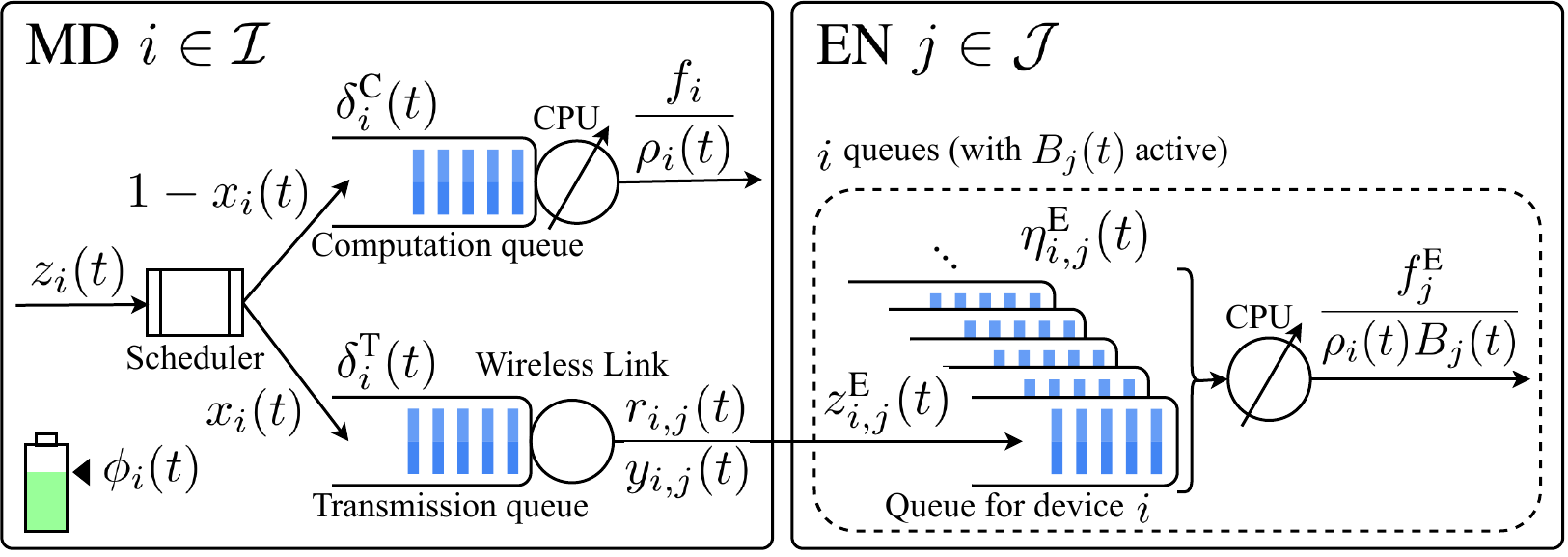}
	\caption{An illustration of MD $i \in \mathcal{I}$ and EN $j \in \mathcal{J}$ in the MEC system.}
	\label{fig1}
\end{figure}

\newcount\LTGbeginlineexample
\newcount\LTGendlineexample 
\newenvironment{ltgexample}%
{\LTGbeginlineexample=\numexpr\inputlineno+1\relax}%
{\LTGendlineexample=\numexpr\inputlineno-1\relax%
	\tcbinputlisting{%
		listing only,
		listing file=\currfilepath, 
		colback=green!5!white,
		colslot=green!25,
		coltitle=black!95,
		coltext=black!05,
		left=8mm,
		title=Corresponding \LaTeX{} code of \texttt{\currfilepath},
		listing options={%
			slot=none,
			language={[LaTeX]TeX},
			escapeinside={},
			firstline=\the\LTGbeginlineexample,
			lastline=\the\LTGendlineexample,
			firstnumber=\the\LTGbeginlineexample,
			basewidth=.5em,
			aboveskip=0mm,
			belowskip=0mm,
			numbers=left,
			xleftmargin=0mm,
			numberstyle=\tiny,
			numbersep=8pt%
		}
	}
}%
We investigate a MEC system consisting of a set of MDs denoted by $\mathcal{I} = \{1, 2, ..., I\}$, along with a set of ENs denoted by $\mathcal{J} = \{1, 2, ..., J\}$, where $I$ and $J$ represent the number of MDs and ENs, respectively. We regard time as a specific episode containing a series of $T$ time slots denoted by $\mathcal{T} = \{1, 2, \ldots, T\}$, each representing a duration of $\tau$ seconds. As shown in Fig.~\ref{fig1}, we consider two separate queues for each MD to organize tasks for local processing or dispatching to ENs, operating in a first-in-first-out (FIFO) manner. The MD's scheduler is responsible for assigning newly arrived tasks to each of the queues at the beginning of the time slot. On the other hand, we assume that each EN $j \in \mathcal{J}$ consists of $I$ FIFO queues, where each queue corresponds to an MD $i \in \mathcal{I}$. When each task arrives at an EN, it is enqueued in the corresponding MD's queue. 

\thispagestyle{empty}
\begin{textblock}{19.1}(1,1.4)
	\vspace{-5mm}
	\noindent \scriptsize \hspace{4mm}
	RAHMATY et al.: QECO: A QOE-ORIENTED COMPUTATION OFFLOADING ALGORITHM BASED ON DEEP REINFORCEMENT LEARNING \hfill 3121
\end{textblock}

We define $z_i(t)$ as the index assigned to the computation task arriving at MD $i \in \mathcal{I}$ in time slot $t \in \mathcal{T}$. Let $\lambda_i(t)$ denote the size of this task in bits. The size of task \( z_i(t) \) is selected randomly and uniformly from a discrete set \( \Lambda = \{\lambda_1, \lambda_2, \ldots, \lambda_{\theta}\} \), where \( \theta \) represents the number of these values. Note that task sizes are drawn from a discrete set since, in many applications, tasks typically come with predefined sizes,~\Cite{wang2020intelligent},~\Cite{zhang2019toward}. We consider $\lambda_i(t) \in \Lambda \cup \{0\}$ to include the case that no task has arrived. We also denote the task's processing density as $\rho_i(t)$ that indicates the number of CPU cycles required to complete the execution of a unit of the task. Furthermore, we denote the deadline of this task by $\Delta_i(t)$ which is the number of time slots that the task must be completed to avoid being dropped.

We define two binary variables, $x_i(t)$ and $y_{i,j}(t)$ for $i \in \mathcal{I}$ and $j \in \mathcal{J}$ to determine the offloading decision and offloading target, respectively. Specifically, $x_i(t)$ indicates whether task $z_i(t)$ is assigned to the computation queue ($x_i(t) = 0$) or to the transmission queue ($x_i(t) = 1$), and $y_{i,j}(t)$ indicates whether task $z_i(t)$ is offloaded to EN $j \in \mathcal{J}$. If the task is dispatched to EN $j$, we set $y_{i,j}(t) = 1$; otherwise, $y_{i,j}(t) = 0$.

\subsection{Communication Model}
 We consider that the tasks in the transmission queue are dispatched to the appropriate ENs via the MD wireless interface. We denote the transmission rate of MD $i$'s interface when communicating with EN $j \in \mathcal{J}$ in time $t$ as $r_{i,j}(t)$. In time slot $t \in \mathcal{T}$, if task $z_i(t)$ is assigned to the transmission queue for computation offloading, we define $l_i^{\text{T}}(t) \in \mathcal{T}$ to represent the time slot when the task is either dispatched to the EN or dropped. We also define $\delta_i^{\text{T}}(t)$ as the number of time slots that task $z_i(t)$ should wait in the queue before transmission. It should be noted that MD $i$ computes the value of $\delta_i^{\text{T}}(t)$ before making a decision. The value of $\delta_i^{\text{T}}(t)$ is computed as follows:
\begin{alignat}{1}
	\delta_i^{\text{T}}(t) =\textcolor{white}{i} \left[ \textcolor{white}{i}\max\limits_{t'\textcolor{white}{i} \in \textcolor{white}{i} \{0,1,\ldots,t-1\}} l_i^{\text{T}}\textcolor{white}{i}(t')-t+1\textcolor{white}{i}\right]^+\textcolor{white}{i},
	\label{1}  
\end{alignat}
where $[\cdot]^+ =$ max$(0, \cdot)$ and $l_i^{\text{T}}(0)=0$ for the simplicity of presentation. Note that the value of $\delta_i^{\text{T}}(t)$ only depends on $l_i^{\text{T}}(t)$ for $t' < t$. If MD $i \in \mathcal{I}$ schedules task $z_i(t)$ for dispatching in time slot $t \in \mathcal{T}$, then it will either be dispatched or dropped in time slot $l_i^{\text{T}}(t)$, which is
\begin{alignat}{1}
	l_i^{\text{T}}(t) = \min \Big\{t + \delta_i^{\text{T}}(t) + \lceil{D_i^{\text{T}}(t)}\rceil - 1, t + \Delta_i(t) - 1\Big\},
	\label{2}  
\end{alignat}
where $D_i^{\text{T}}(t)$ refers to the number of time slots required for the transmission of task $z_i(t)$ from MD $i \in \mathcal{I}$ to EN $j \in \mathcal{J}$. We have
\begin{alignat}{1}
	D_i^{\text{T}}(t) =  \sum_{j \in \mathcal{J}} y_{i,j}(t) {\lambda_i(t) \over r_{i,j}(t)\tau}.
	\label{3}  
\end{alignat}
Let $E_i^{\text{T}}(t)$ denote the energy consumption of the transmission from MD $i \in \mathcal{I}$ to EN $j \in \mathcal{J}$. We have
\begin{alignat}{1}
	E_i^{\text{T}}(t) = D_i^{\text{T}}(t)p_i^{\text{T}}(t)\tau,
	\label{4}  
\end{alignat}
where $p_i^{\text{T}}(t)$ represents the power consumption of the communication link of MD $i \in \mathcal{I}$ in time slot $t$.
\subsection{Computation Model}
The computation tasks can be executed either locally on the MD or on the EN. In this subsection, we provide a detailed explanation of these two cases.
\subsubsection{Local Execution}
We model the local execution by a queuing system consisting the computation queue and the MD processor. Let $f_i$ denote the MD $i$'s processing power (in cycle per second). When task $z_i(t)$ is assigned to the computation queue at the beginning of time slot $t \in \mathcal{T}$, we define $l_i^{\text{C}}(t) \in \mathcal{T}$ as the time slot during which task $z_i(t)$ will either be processed or dropped. If the computation queue is empty, $l_i^{\text{C}}(t) = 0$. Let $\delta_i^{\text{C}}(t)$ denote the number of remaining time slots before processing task $z_i(t)$ in the computation queue. We have:
\begin{alignat}{1}
	\delta_i^{\text{C}}(t) = \left[ \max \limits_{t' \in \{0,1,\ldots,t-1\}} l_i^{\text{C}}(t')-t+1 \right]^+.
	\label{5}  
\end{alignat}
In the equation above, the term $\max_{t' \in \{0, 1, \ldots, t-1\}} l_i^{\text{C}}(t')$ denotes the time slot at which each existing task in the computation queue, which arrived before time slot $t$, is either processed or dropped. Consequently, $\delta_i^{\text{C}}(t)$ denotes the number of time slots that task $z_i(t)$ should wait before being processed. We denote the time slot in which task $z_i(t)$ will be completely processed by $l_i^{\text{C}}(t)$ if it is assigned to the computation queue for local processing in time slot $t$. We have
\begin{alignat}{1}
	l_i^{\text{C}}(t) = \min \bigg\{t + \delta_i^{\text{C}}(t) + \lceil D_i^{\text{C}}(t) \rceil - 1, t + \Delta_i(t) - 1\bigg\}.
	\label{6}  
\end{alignat}
The task $z_i(t)$ will be immediately dropped if its processing is not completed by the end of the time slot $t + \Delta_i(t) - 1$. In addition, we introduce $D_i^{\text{C}}(t)$ as the number of time slots required to complete the processing of task $z_i(t)$ on MD $i \in \mathcal{I}$. 
\begin{alignat}{1}
	D_i^{\text{C}}(t) = { \lambda_i(t)  \over  f_i  \tau /  \rho_i(t)}.
	\label{7}  
\end{alignat}

\thispagestyle{empty}
\begin{textblock}{19.1}(1,1.4)
	\vspace{-5mm}
	\noindent \scriptsize \hspace{4mm}
	3122 \hfill IEEE TRANSACTIONS ON NETWORK SCIENCE AND ENGINEERING, VOL. 12, NO. 4, JULY/AUGUST 2025
\end{textblock}

To compute the MD's energy consumption in the time slot $t \in \mathcal{T}$, we define $E_i^{\text{L}}(t)$ as:
\begin{alignat}{1}
	E_i^{\text{L}}(t) =  D_i^{\text{C}}(t) p_i^{\text{C}}  \tau, 
	\label{8}  
\end{alignat}
where $p_i^{\text{C}} = 10^{-27}(f_i)^3$ represents the energy consumption of MD $i$'s CPU frequency \Cite{mao2016dynamic}.
\subsubsection{Edge Execution}
We model the edge execution by the queues associated with MDs deployed at ENs. If computation task $z_i(t')$ is dispatched to EN $j$ in time $t' < t$, we let $z_{i,j}^{\text{E}}(t)$, $\lambda_{i,j}^{\text{E}}(t)$ (in bits), and $\rho_{i,j}^{\text{E}}(t)$ denote the unique index of the task, the task size, and the number of CPU cycles required per unit of the task in the $i^{\text{th}}$ queue at EN $j$, respectively. We define $\eta_{i,j}^{\text{E}}(t)$ (in bits) as the length of this queue at the end of time slot $t \in \mathcal{T}$. We refer to a queue as an active queue in a certain time slot if it is not empty. That being said, if at least one task is already in the queue from previous time slots or there is a task arriving at the queue, that queue is active. We define $\mathcal{B}_j(t)$ to denote the set of active queues at EN $j$ in time slot $t$.
\begin{alignat}{1}
	\mathcal{B}_j(t) = \textcolor{white}{i}\bigg\{i \textcolor{white}{i}\big|\, i \in \mathcal{I}, \lambda_{i,j}^{\text{E}}(t)>0\,\, \textcolor{white}{i}\text{or}\textcolor{white}{i} \,\, \eta_{i,j}^{\text{E}}(t-1)>0\textcolor{white}{i}\bigg\}.\textcolor{white}{i}
	\label{9}  
\end{alignat} 

We introduce $b_j(t) \triangleq |\mathcal{B}_j(t)|$ that represents the number of active queues in EN $j \in \mathcal{J}$ in time slot $t \in \mathcal{T}$. In each time slot $t \in \mathcal{T}$, the EN's processing power is divided among its active queues using a generalized processor sharing method~\Cite{parekh1993generalized}. Let variable $f_j^{\text{E}}$ (in cycles per second) represent the computational capacity of EN $j$. Therefore, EN $j$ can allocate computational capacity of $f_j^{\text{E}}/(\rho_i(t) b_j(t))$ to each MD $i \in \mathcal{B}_j(t)$ during time slot $t$. To calculate the length of the computation queue for MD $i \in \mathcal{I}$ in EN $j \in \mathcal{J}$, we define $\omega_{i,j}(t)$ (in bits) to represent the number of bits from dropped tasks in that queue at the end of time slot $t \in  \mathcal{T}$. The backlog of the queue, referred to as $\eta_{i,j}^{\text{E}}(t)$ is given~by:
\begin{alignat}{1}
	\eta_{i,j}^{\text{E}}(t)\hspace{-0.8mm}=\hspace{-1mm}\left[\eta_{i,j}^{\text{E}}(t-1)\hspace{-0.6mm}+\hspace{-0.7mm}\lambda_{i,j}^{\text{E}}(t)\hspace{-0.7mm}-\hspace{-0.7mm}{f_j^{\text{E}}\tau\over \rho_{i,j}^{\text{E}}(t)b_j(t)} -\omega_{i,j}(t)\right]^+\hspace{-1mm}.
	\label{10}  
\end{alignat}
We also define $l_{i,j}^{\text{E}}(t) \in \mathcal{T}$ as the time slot during which the offloaded task $z_{i,j}^{\text{\text{E}}}(t)$ is either processed or dropped by EN $j$. Given the uncertain workload ahead at EN $j$, neither MD $i$ nor EN $j$ has information about $l_{i,j}^{\text{E}}(t)$ until the corresponding task $z_{i,j}^{\text{E}}(t)$ is either processed or dropped. Let $\hat{l}_{i,j}^{\text{E}}(t)$ represent the time slot at which the execution of task $z_{i,j}^{\text{E}}(t)$ starts. In mathematical terms, for $i \in \mathcal{I}$, $j \in \mathcal{J}$, and $t \in \mathcal{T}$, we have:
\begin{alignat}{1}
	\hat{l}_{i,j}^{\text{E}}(t) = \max \bigg\{t, \max \limits_{t' \in \{0,1,\ldots,t-1\}} l_{i,j}^{\text{E}}(t')+1\bigg\},
	\label{11}  
\end{alignat}
where $l_{i,j}^{\text{E}}(0) = 0$. Indeed, the initial processing time slot of task $z_{i,j}^{\text{E}}(t)$ at EN should not precede the time slot when the task was enqueued or when the previously arrived tasks were processed or dropped. Therefore, $l_{i,j}^{\text{E}}(t)$ is the time slot that satisfies the following constraints. 
\begin{alignat}{1}
	\sum_{t'=\hat{l}_{i,j}^{\text{E}}(t)}^{l_{i,j}^{\text{E}}(t)}{f_j^{\text{E}}\tau \over \rho_{i,j}^{\text{E}}(t)b_j(t')}\mathbbm{1}(i \in \mathcal{B}_j(t'))  \geq   \lambda_{i,j}^{\text{E}}(t),
	\label{12}  \\
	\sum_{t'=\hat{l}_{i,j}^{\text{E}}(t)}^{l_{i,j}^{\text{E}}(t)-1}{f_j^{\text{E}}\tau \over \rho_{i,j}^{\text{E}}(t)b_j(t')}\mathbbm{1}(i \in \mathcal{B}_j(t')) < \lambda_{i,j}^{\text{E}}(t),
	\label{13}  
\end{alignat}
where $\mathbbm{1} (z \in \mathbb{Z})$ is the indicator function. In particular, the total processing capacity that EN $j$ allocates to MD $i$ from the time slot $\hat{l}_{i,j}^{\text{E}}(t)$ to the time slot $l_{i,j}^{\text{E}}(t)$ should exceed the size of task $z_{i,j}^{\text{E}}(t)$. Conversely, the total allocated processing capacity from the time slot $l_{i,j}^{\text{E}}(t)$ to the time slot $l_{i,j}^{\text{E}}(t)-1$ should be less than the task's size.

Additionally, we define $D_{i,j}^{\text{E}}(t)$ to represent the quantity of processing time slots allocated to task $z_{i,j}^{\text{E}}(t)$ when executed at EN $j$. This value is given by:
\begin{alignat}{1}
		\textcolor{black}{D_{i,j}^{\text{E}}(t) =  l_{i,j}^{\text{E}}(t)-\hat{l}_{i,j}^{\text{E}}(t).}
	\label{14}  
\end{alignat}

We also define $E_{i,j}^{\text{E}}(t)$ as the energy consumption of processing at EN $j$ in time slot $t$ by MD $i$. This can be calculated~as:
\begin{alignat}{1}
	\textcolor{black}{E_{i,j}^{\text{E}}(t) =  \sum_{t'=\hat{l}_{i,j}^{\text{E}}(t)}^{l_{i,j}^{\text{E}}(t)}\hspace{-2mm}{\, p_j^{\text{E}} \tau \over \,\, b_j(t')\,\,}\mathbbm{1}(i \in \mathcal{B}_j(t')),}  
	\label{15}  
\end{alignat}
where $p_j^{\text{E}}$ is constant and denotes the energy consumption of the EN $j$'s processor when operating at full capacity. 

In addition to the energy consumed by EN $j$ for task processing, we also take into account the energy consumed by the MD $i$'s user interface in the standby state while waiting for task completion at the EN $j$. We define $E_{i}^{\text{I}}(t)$ as the energy consumption associated with the user interface of MD $i \in \mathcal{I}$, which is given by
\begin{alignat}{1}
		\textcolor{black}{E_{i}^{\text{I}}(t) =  \sum_{j \in \mathcal{J}}y_{i,j}(t) D_{i,j}^{\text{E}}(t) p_i^{\text{I}} \tau, }
	\label{16}
\end{alignat}
where $p_i^{\text{I}}$ is the standby energy consumption of MD $i \in \mathcal{I}$.  
\textcolor{black}{Recall that $y_{i,j}(t)$ is the binary offloading indicator. Moreover, among all $j \in \mathcal{J}$, $y_{i,j}=1$ only for one $j$ which is the corresponding~EN.}


\section{Task Offloading problem Formulation}
\label{section:IV}

Based on the introduced system model, we present the computation task offloading problem in this section. Our primary goal is to enhance each MD's QoE individually by taking the dynamic demands of MDs into account. To achieve this, we approach the optimization problem as an MDP, aiming to maximize the MD's QoE by striking a balance among key QoE factors, including task completion, task delay, and energy consumption. To prioritize QoE factors, we utilize the MD's energy level of the battery, which plays a crucial role in decision-making. Specifically, when an MD observes its state (e.g. task size, queue details, and battery status) and encounters a newly arrived task, it selects an appropriate action for that task. The selected action, based on the observed state, will result in enhanced QoE. Each MD strives to maximize its long-term QoE by optimizing the policy mapping from states to actions. In what follows, we first present the state space, action space, and QoE function, respectively. We then formulate the QoE maximization problem for each MD.

\thispagestyle{empty}
\begin{textblock}{19.1}(1,1.4)
	\vspace{-5mm}
	\noindent \scriptsize \hspace{4mm}
	RAHMATY et al.: QECO: A QOE-ORIENTED COMPUTATION OFFLOADING ALGORITHM BASED ON DEEP REINFORCEMENT LEARNING \hfill 3123
\end{textblock}

\subsection{State Space}
A state in our MDP represents a conceptual space that comprehensively describes the state of an MD facing the~environment. We represent the MD $i$'s state in time slot $t$ as vector $\boldsymbol{s}_i(t)$ that includes the newly arrived task size, the queues information, the MD's energy level of the battery, and the workload history at the ENs. The MD observes this vector at the beginning of each time slot. The vector $\boldsymbol{s}_i(t)$ is defined as follows:
\begin{alignat}{1}
\boldsymbol{s}_i(t) = \Big(\lambda_i(t), \delta_i^{\text{C}}(t), \delta_i^{\text{T}}(t), \boldsymbol{\eta}_i^{\text{E}}(t-1),\phi_i(t), \mathcal{H}(t) \Big),
	\label{18}
\end{alignat}
where vector $\boldsymbol{\eta}_i^{\text{E}}(t-1) = (\eta_{i,j}^{\text{E}}(t-1))_{j \in \mathcal{J}}$ represents the queues length of MD $i$ in ENs at the previous time slot and is computed by the MD according to \eqref{10}. 
\textcolor{black}{Since MDs are assumed to be battery-operated, their operation modes are aligned with those of real-world devices. Let $\phi_i(t)$ represent the \textcolor{black}{battery level percentage of MD $i$} at time slot $t$, where $\phi_i(t)$ is selected from the discrete set $\Phi = \{\phi_1, \phi_2, \phi_3\}$, corresponding to ultra power-saving, power-saving, and performance modes, respectively.}

In addition, to predict future EN workloads, we define the matrix $\mathcal{H}(t)$ as historical data, indicating the number of active queues for all ENs. This data is recorded over $T^{\text{s}}$ time slots, ranging from $t-T^{\text{s}}$ to $t-1$, in $T^{\text{s}} \times J$ matrix. For EN $j$ workload history at $i^{th}$ time slot from $T^{\text{s}}-t$, we define $h_{i,j}(t)$ as:
\begin{alignat}{1}
	h_{i,j}(t) = b_j(t - T^{\text{s}} + i - 1).
	\label{19}
\end{alignat}
EN $j \in \mathcal{J}$ broadcasts $b_j(t)$ at the end of each time slot. 

We define vector $\mathcal{S}$ as the discrete and finite state space for each MD. The size of the set $\mathcal{S}$ is given by $\Lambda \times T^2 \times \mathcal{U} \times 3 \times I^{T^{\text{s}} \times J}$, where $\mathcal{U}$ is the set of available queue length values at an EN over $T$ time slots.




\subsection{Action Space}
The action space represents the agent's decisions. We define $\boldsymbol{a}_i(t)$ to denote the action taken by MD $i \in \mathcal{I}$ in time slot $t \in \mathcal{T}$. These actions involve two decisions, (a) Offloading decision to determine whether or not to offload the task, and (b) Offloading target to determine the EN to send the offloaded tasks. Thus, the action of MD $i$ in time slot $t$ can be concisely expressed as the following action tuple: 
\begin{alignat}{1}
	\boldsymbol{a}_i(t) = (x_i(t), \boldsymbol{y}_i(t)),
	\label{20}
\end{alignat}
where vector $\boldsymbol{y}_i(t)=(y_{i,j}(t))_{j \in \mathcal{J}}$ represents the selected EN for offloading this task. In Section~\ref{section:1}, we will discuss the size of this action space.


\subsection{QoE Function}
\textcolor{black}{The QoE function reflects user satisfaction with task computation, whether using local or edge resources. QoE accounts for task completion rate, processing delay, and energy consumption. Each user may have unique QoE requirements~\Cite{lu2020edge}. Therefore, we define a multi-dimensional, adaptive structure to assess QoE, balancing these factors according to the MDs preferences, such as prioritizing reduced task delay or energy saving, which may vary over time. To capture personalized QoE requirements, we define QoE as a weighted sum of the above factors. Each MD dynamically adjusts these weights to reflect the importance of different factors based on its energy modes, which are performance, power-saving, and ultra-power-saving modes. We now calculate task delay and energy consumption and then introduce the associated cost and QoE function.}


	
 Given the selected action $\boldsymbol{a}_i(t)$ in the observed state $\boldsymbol{s}_i(t)$, we represent $\mathcal{D}_i(\boldsymbol{s}_i(t), \boldsymbol{a}_i(t))$ as the delay of task $z_i(t),$ which indicates the number of time slots from time slot $t$ to the time slot in which task $z_i(t)$ is processed. It is calculated by: \vspace{3mm}

$\mathcal{D}_i(\boldsymbol{s}_i(t),\boldsymbol{a}_i(t)) = (1-x_i(t))\Big(l_i^{\text{C}}(t) - t + 1\Big) $
\begin{alignat}{1}
	\hspace{0.3cm} +\, x_i(t) \Bigg(  \sum\limits_{j \in \mathcal{J}} \sum\limits_{t'=t}^{T} \mathbbm{1}\big(z_{i,j}^{\text{E}}(t') = z_i(t)\big)\, l_{i,j}^{\text{E}}(t') - t +1\Bigg),
	\label{21}  
\end{alignat} 
where $\mathcal{D}_i(\boldsymbol{s}_i(t),\boldsymbol{a}_i(t))= 0$ if task $z_i(t)$ is dropped. Correspondingly, we denote the energy consumption of task $z_i(t)$ when taking action $\boldsymbol{a}_i(t)$ in the observed state $\boldsymbol{s}_i(t)$ as $\mathcal{E}_i(\boldsymbol{s}_i(t),\boldsymbol{a}_i(t))$, which is \vspace{2.2mm}

\textcolor{black}{$\mathcal{E}_i(\boldsymbol{s}_i(t),\boldsymbol{a}_i(t)) = (1-x_i(t)) E_i^{\text{L}}(t)+x_i(t) \bigg( E_i^{\text{T}}(t) + $\vspace{0mm}
\begin{alignat}{1}
	\hspace{1.2cm}    E_i^{\text{I}}(t) +  \sum\limits_{j \in \mathcal{J}}  \sum\limits_{t'=t}^{T}\mathbbm{1}\big(z_{i,j}^{\text{E}}(t') = z_i(t)\big)
	  E_{i,j}^{\text{E}}(t')\bigg).
	\label{22}  
\end{alignat}}

\textcolor{black}{To define associated cost, we use a weighted sum of task delay $\mathcal{D}_i(\boldsymbol{s}_i(t),\boldsymbol{a}_i(t))$ and energy consumption $\mathcal{E}_i(\boldsymbol{s}_i(t),\boldsymbol{a}_i(t))$, where the MD dynamically adjusts the weights based on its energy level to reflect the preference for each factor.}
Given the delay and energy consumtion of task $z_i(t)$, we also define $\mathcal{C}_i(\boldsymbol{s}_i(t),\boldsymbol{a}_i(t))$ that denotes the assosiate cost of task $z_i(t)$ given the action $\boldsymbol{a}_i(t)$ in the state $\boldsymbol{s}_i(t)$. \vspace{2.2mm}

$\mathcal{C}_i(\boldsymbol{s}_i(t),\boldsymbol{a}_i(t)) =$
\begin{alignat}{1}
	\phi_i(t) \, \mathcal{D}_i(\boldsymbol{s}_i(t),\boldsymbol{a}_i(t)) +(1-\phi_i(t)) \, \mathcal{E}_i(\boldsymbol{s}_i(t),\boldsymbol{a}_i(t)),
	\label{23}  
\end{alignat}
where $\phi_i(t)$ represents the MD $i$'s \textcolor{black}{energy level}. When the MD is operating in performance mode, the primary focus is on minimizing task delays, thus the delay contributes more to the cost. On the other hand, when the MD switches to ultra power-saving mode, the main attention is directed toward reducing power consumption.

Finally, we define $\boldsymbol{q}_i(\boldsymbol{s}_i(t),\boldsymbol{a}_i(t))$ as the QoE associated with task $z_i(t)$ given the selected action $\boldsymbol{a}_i(t)$ and the observed state $\boldsymbol{s}_i(t)$. The QoE function is defined as follows $\boldsymbol{q}_i(\boldsymbol{s}_i(t),\boldsymbol{a}_i(t)) =$
\begin{alignat}{1}
	\hspace*{8mm}
	\begin{cases} 
		\mathcal{R} - \mathcal{C}_i(\boldsymbol{s}_i(t),\boldsymbol{a}_i(t)) & \hspace*{-2mm} \text{if task $z_i(t)$ is processed,} \\
		- \hspace*{0.8mm} \mathcal{E}_i(\boldsymbol{s}_i(t),\boldsymbol{a}_i(t)) &	\hspace*{-2mm} \text{if task $z_i(t)$ is dropped,}
	\end{cases}
	\label{26}  
\end{alignat}
where $\mathcal{R} > 0$ represents a constant reward for task completion. If $z_i(t) = 0$, then $\boldsymbol{q}_i(\boldsymbol{s}_i(t), \boldsymbol{a}_i(t)) = 0$. Throughout the rest of this paper, we adopt the shortened notation $\boldsymbol{q}_i(t)$ to represent $\boldsymbol{q}_i(\boldsymbol{s}_i(t), \boldsymbol{a}_i(t))$.

\subsection{Problem Formulation}
We define the task offloading policy for MD $i \in \mathcal{I}$ as a mapping from its state to its corresponding action, denoted by i.e., $\pi_i : \mathcal{S} \rightarrow \mathcal{A}$. Especially, MD $i$ determines an action $\boldsymbol{a}_i(t) \in \mathcal{A}$, according to policy $\pi_i$ given the observed environment state $\boldsymbol{s}_i(t) \in \mathcal{S}$. The MD aims to find its optimal policy $\pi_i^*$ which maximizes the long-term QoE,
\begin{alignat}{1}
	\pi_i^* = \text{arg}\,\,  \max\limits_{\pi_i}  \mathbbm{E} \Bigg[ \sum\limits_{t \in \mathcal{T}}  \gamma^{t-1}\boldsymbol{q}_i(t) \Bigg| \pi_i \Bigg],
	\label{24}  
\end{alignat}
where $\gamma \in (0,1]$ is a discount factor and determines the balance between instant QoE and long-term QoE. As $\gamma$ approaches 0, the MD prioritizes QoE within the current time slot exclusively. Conversely, as $\gamma$ approaches 1, the MD increasingly factors in the cumulative long-term QoE. The expectation $\mathbb{E}[\cdot]$ is taken into consideration of the time-varying system environments. Solving the optimization problem in \eqref{24} is particularly challenging due to the dynamic nature of the network. To address this challenge, we introduce a DRL-based offloading algorithm to learn the mapping between each state-action pair and their long-term QoE.
\section{DRL-Based Offloading Algorithm} \label{section:V}
We now present QECO algorithm so as to address the distributed offloading decision-making of MDs. The aim is to empower MDs to identify the most efficient action that maximizes their long-term QoE. In the following, we introduce a neural network that characterizes the MD's state-action Q-values mapping, followed by a description of the information exchange between the MDs and ENs.

\subsection{DQN-based Approach}
\begin{figure}
	\centering
	\includegraphics[width=1\linewidth]{ 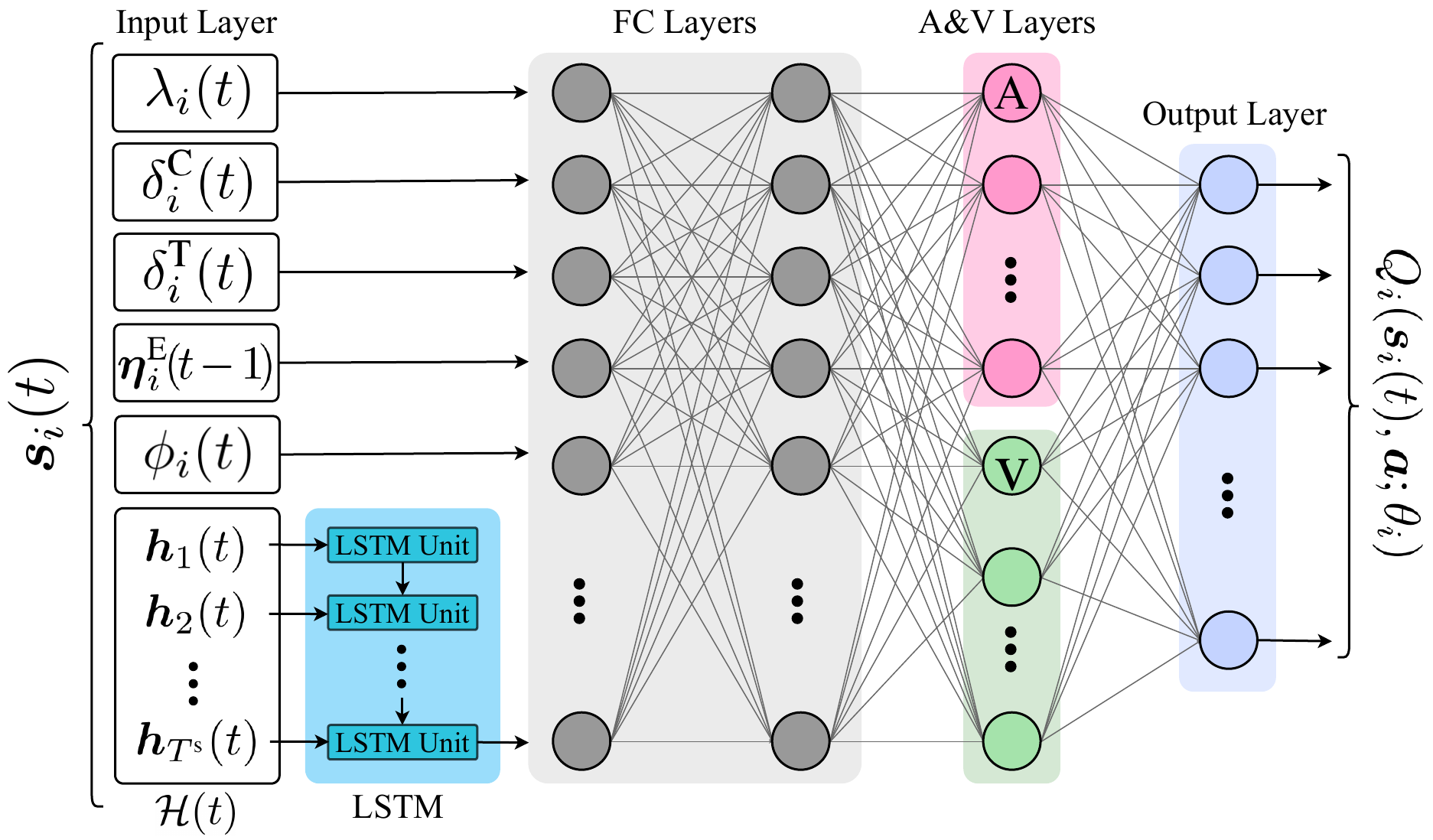}
	\captionsetup{name=Fig.}
	\vspace*{-3mm}
	\caption{The neural network of MD $i \in \mathcal{I}$, which characterize the Q-value of each action $\boldsymbol{a} \in \mathcal{A}$ under state $\boldsymbol{s}_i(t) \in \mathcal{S}$.}
	\label{DQN}
\end{figure}

We utilize the DQN technique to find the mapping between each state-action pair to Q-values in the formulated MDP. As shown in Fig.~\ref{DQN}, each MD $i \in \mathcal{I}$ is equipped with a neural network comprising six layers. These layers include an input layer, an LSTM layer, two dense layers, an advantage-value (A\&V) layer, and an output layer. The parameter vector $\theta_i$ of MD $i$'s neural network is defined to maintain the connection weights and neuron biases across all layers. For MD $i \in \mathcal{I}$, we utilize the state information as the input of neural network. The state information $\lambda_i(t)$, $\delta_i^{\text{C}}(t)$, $\delta_i^{\text{T}}(t)$, $\phi_i(t)$, and $\boldsymbol{\eta}_i^{\text{E}}(t-1)$ are directly passed to the dense layer, while the state information $\mathcal{H}(t)$ is first supplied to the LSTM layer and then the resulting output is sent to the dense layer. The roles and responsibilities of each layer are detailed as follows.

\thispagestyle{empty}
\begin{textblock}{19.1}(1,1.4)
	\vspace{-5mm}
	\noindent \scriptsize \hspace{4mm}
	3124 \hfill IEEE TRANSACTIONS ON NETWORK SCIENCE AND ENGINEERING, VOL. 12, NO. 4, JULY/AUGUST 2025
\end{textblock}

\subsubsection{Predicting Workloads at ENs}
In order to capture the dynamic behavior of workloads at the ENs, we employ the LSTM network \Cite{hochreiter1997long}. This network maintains a memory state $\mathcal{H}(t)$ that evolves over time, enabling the neural network to predict future workloads at the ENs based on historical data. By taking the matrix $\mathcal{H}(t)$ as an input, the LSTM network learns the patterns of workload dynamics. The architecture of the LSTM consists of $T^{\text{s}}$ units, each equipped with a set of hidden neurons, and it processes individual rows of the matrix $\mathcal{H}(t)$ sequentially. Through this interconnected design, MD tracks the variations in sequences from $\boldsymbol{h}_1(t)$ to $\boldsymbol{h}_{T^{\text{s}}}(t)$, where vector $\boldsymbol{h}_i(t) = (h_{i,j}(t))_{j \in \mathcal{J}}$, thereby revealing workload fluctuations at the ENs across different time slots. The final LSTM unit produces an output that encapsulates the anticipated workload dynamics, and is then connected to the subsequent layer neurons for further learning.
\subsubsection{State-Action Q-Value Mapping}
The pair of dual dense layers plays a crucial role in learning the mapping of Q-values from the current state and the learned load dynamics to the corresponding actions. The dense layers consist of a cluster of neurons that employ rectified linear units (ReLUs) as their activation functions. In the initial dense layer, connections are established from the neurons in the input layer and the LSTM layer to each neuron in the dense layer. The resulting output of a neuron in the dense layer is connected to each neuron in the subsequent dense layer. In the second layer, the outputs from each neuron establish connections with all neurons in the A\&V layers.

\subsubsection{Dueling-DQN Approach for Q-Value Estimation}
In the neural network architecture, the A\&V layer and the output layer incorporate the principles of the dueling-DQN \Cite{wang2016dueling} to compute action Q-values. The fundamental concept of dueling-DQN involves two separate learning components: one for action-advantage values and another for state-value. This approach enhances Q-value estimation by separately evaluating the long-term QoE attributed to states and actions.

\thispagestyle{empty}
\begin{textblock}{19.1}(1,1.4)
	\vspace{-5mm}
	\noindent \scriptsize \hspace{4mm}
	RAHMATY et al.: QECO: A QOE-ORIENTED COMPUTATION OFFLOADING ALGORITHM BASED ON DEEP REINFORCEMENT LEARNING \hfill 3125
\end{textblock}

The A\&V layer consists of two distinct dense networks referred to as network A and network V. Network A's role is to learn the action-advantage value for each action, while network V focuses on learning the state-value. For an MD $i \in \mathcal{I}$, we define $V_i(\boldsymbol{s}_i(t); \theta_i)$ and $A_i(\boldsymbol{s}_i(t), \boldsymbol{a}; \theta_i)$ to denote the state-value and the action-advantage value of action $\boldsymbol{a} \in \mathcal{A}$ under state $\boldsymbol{s}_i(t) \in \mathcal{S}$, respectively. The parameter $\theta_i$ is responsible for determining these values, and it can be adjusted when training the QECO algorithm.

For an MD $i \in \mathcal{I}$, the A\&V layer and the output layer collectively determine $Q_i(\boldsymbol{s}_i(t), \boldsymbol{a}; \theta_i)$, representing the resulting Q-value under action $\boldsymbol{a} \in \mathcal{A}$ and state $\boldsymbol{s}_i(t) \in \mathcal{S}$, as follows: \vspace{4mm}

$Q_i(\boldsymbol{s}_i(t), \boldsymbol{a}; \theta_i) = V_i(\boldsymbol{s}_i(t);\theta_i) + $
\begin{alignat}{1}
	\hspace{0.74cm}  \Bigg( A_i(\boldsymbol{s}_i(t),\boldsymbol{a};\theta_i) - {1 \over |\mathcal{A}|} \sum\limits_{\boldsymbol{a}' \in \mathcal{A}}(A_i(\boldsymbol{s}_i(t),\boldsymbol{a}';\theta_i) \Bigg),
	\label{25}  
\end{alignat}
where $\theta_i$ establishes a functional relationship that maps Q-values to pairs of state-action.


\subsection{QoE-Oriented DRL-Based Algorithm}
\label{section:1}


The QECO algorithm is meticulously designed to optimize the allocation of computational tasks between MDs and ENs. Since the training of neural networks imposes an extensive computational workload on MDs, we enable MDs to utilize ENs for training their neural networks, effectively reducing their computational workload. For each MD $i \in \mathcal{I}$, there is an associated EN, denoted as EN $j_i \in \mathcal{J}$, which assists in the training process. This EN possesses the highest transmission capacity among all ENs. We define $\mathcal{I}_j \subset \mathcal{I}$ as the set of MDs for which training is executed by EN $j \in \mathcal{J}$, i.e. $\mathcal{I}_j = \{i \in \mathcal{I} | j_i = j\}$. This approach is feasible due to the minimal information exchange and processing requirements for training compared to MD's tasks. The algorithms to be executed at MD $i \in \mathcal{I}$ and EN $j \in \mathcal{J}$ are given in Algorithms ~\ref{alg:cap} and ~\ref{alg:cap2}, respectively. The core concept involves training neural networks with MD experiences (i.e., state, action, QoE, next state) to map Q-values to each state-action pair. This mapping allows MD to identify the action in the observed state with the highest Q-value and maximize its long-term QoE.

In detail, EN $j \in \mathcal{J}$ maintains a replay buffer denotes as $\mathcal{M}_i$ with two neural networks for MD $i$: $\textit{Net}_i^{\text{E}}$, denoting the evaluation network, and $\textit{Net}_i^{\text{T}}$, denoting the target network, which have the same neural network architecture. However, they possess distinct parameter vectors $\theta^{\text{E}}_i$ and $\theta^{\text{T}}_i$, respectively. Their Q-values are represented by $Q_i^{\text{E}}(\boldsymbol{s}_i(t), \boldsymbol{a}; \theta^{\text{E}}_i)$ and $Q_i^{\text{T}}(\boldsymbol{s}_i(t), \boldsymbol{a}; \theta^{\text{T}}_i)$ for MD $i \in \mathcal{I}_j$, respectively, associating the action $\boldsymbol{a} \in \mathcal{A}$ under the state $\boldsymbol{s}_i(t) \in \mathcal{S}$. The replay buffer records the observed experience $(\boldsymbol{s}_i(t), \boldsymbol{a}_i(t), \boldsymbol{q}_i(t), \boldsymbol{s}_i(t+1))$ of MD $i$. Moreover, $\textit{Net}_i^{\text{E}}$ is responsible for action selection, while $\textit{Net}_i^{\text{T}}$ characterizes the target Q-values, which represent the estimated long-term QoE resulting from an action in the observed state. The target Q-value serves as the reference for updating the network parameter vector $\theta^{\text{E}}_i$. This update occurs through the minimization of disparities between the Q-values under $\textit{Net}_i^{\text{E}}$ and $\textit{Net}_i^{\text{T}}$. In the following, we introduce the offloading decision algorithm of MD $i \in \mathcal{I}$ and the training process algorithm running in EN $j \in \mathcal{J}$.

\begin{algorithm}[tbp]
	\caption{\textcolor{black}{Offloading Decision Algorithm at MD $i \in \mathcal{I}$}}\label{alg:cap}
	\begin{algorithmic}[1]
		\renewcommand{\algorithmicrequire}{\textbf{\textcolor{black}{Input:}}} 
		\renewcommand{\algorithmicensure}{\textbf{\textcolor{black}{Output:}}}
		\Require \textcolor{black}{ state space $\mathcal{S}$, action space $\mathcal{A}$, parameters vector~$\theta_i^{\text{E}}$}
		\Ensure \textcolor{black}{MD $i \in \mathcal{I}$ experience  \textcolor{black}{ $(\hspace{-0.2mm}\boldsymbol{s}_i(\hspace{-0.2mm}t\hspace{-0.2mm}), \boldsymbol{a}_i(\hspace{-0.2mm}t\hspace{-0.2mm}), \boldsymbol{q}_i(\hspace{-0.2mm}t\hspace{-0.2mm}), \boldsymbol{s}_i(\hspace{-0.2mm}t\hspace{-0.3mm}+\hspace{-0.3mm}\hspace{-0.2mm}1\hspace{-0.2mm}\hspace{-0.2mm})\hspace{-0.2mm})$}}
		\For {episode 1 to $N^{\text{ep}}$}
		\State Initialize $\boldsymbol{s}_i(1)$
		\For {time slot $t \in \mathcal{T}$}
		\If{MD $i$ receives a new task $z_i(t)$}
		\State Send an \textit{UpdateRequest} to EN $j_i$;
		\State Receive network parameter vector $\theta_i^{\text{E}}$;
		\State Select action $\boldsymbol{a}_i(t)$ based on \eqref{26};
		\EndIf
		\State Observe a set of QoEs $\{\boldsymbol{q}_i(t'), t' \in \mathcal{F}_i^t\}$;
		\State Observe\textcolor{white}{i}the next\textcolor{white}{i}state $\textbf{\textit{s}}_i(t+1)$;\textcolor{white}{i}
		\For {each task $z_i(t')$ where $t' \in \mathcal{F}_i^t$} 
		\State Send \hspace{-1mm} $(\boldsymbol{s}_i(t'), \boldsymbol{a}_i(t'), \boldsymbol{q}_i(t'), \boldsymbol{s}_i(t'\hspace{-1mm}+1))$ to EN $j_i$;
		\EndFor
		\EndFor
		\EndFor
	\end{algorithmic}
\end{algorithm}

\subsubsection{Offloading Decision Algorithm at MD $i \in \mathcal{I}$}
 We analyze a series of episodes, where $N^{\text{ep}}$ denotes the number of them. At the beginning of each episode, if MD $i \in \mathcal{I}$ receives a new task $z_i(t)$, it initializes the state $\mathcal{S}_i(1)$ and sends an $\textit{UpdateRequest}$ to EN $j_i$. After receiving the requested vector $\theta_i^{\text{E}}$ of $\textit{Net}_i^{\text{E}}$ from EN $j_i$, MD $i$ chooses the following action for task $z_i(t)$.
\begin{alignat}{1}
	\hspace*{-2mm}\boldsymbol{a}_i(t) \hspace*{-0.5mm} = \hspace*{-0.5mm}
	\begin{cases} 
		\text{arg $\max_{\boldsymbol{a}\in \mathcal{A}}Q_i^{\text{E}}(\boldsymbol{s}_i(t), \boldsymbol{a}; \theta^{\text{E}}_i)$,} & \text{w.p. $1-\boldsymbol{\epsilon}$,} \\
 	\text{pick a random action from $\mathcal{A}$,} & \text{w.p. $\boldsymbol{\epsilon}$,}
	\end{cases}
	\label{26}  
\end{alignat}
where w.p. stands for with probability, and $\boldsymbol{\epsilon}$ represents the random exploration probability. The value of $Q_i^{\text{E}}(\boldsymbol{s}_i(t), \boldsymbol{a}; \theta^{\text{E}}_i)$ indicates the Q-value under the parameter $\theta^{\text{E}}_i$ of the neural network $\textit{Net}_i^{\text{E}}$. Specifically, the MD with a probability of $1 - \boldsymbol{\epsilon}$ selects the action associated with the highest Q-value under $\textit{Net}_i^{\text{E}}$ in the observed state $\boldsymbol{s}_i(t)$.

In the next time slot $t+1$, MD $i$ observes the state $\mathcal{S}_i(t+1)$. However, due to the potential for tasks to extend across multiple time slots, QoE $\boldsymbol{q}_i(t)$ associated with task $z_i(t)$ may not be observable in time slot $t+1$. On the other hand, MD $i$ may observe a group of QoEs associated with some tasks $z_i(t')$ in time slots $t' \leq t$. For each MD $i$, we define the set $\mathcal{F}_i^t \subset \mathcal{T}$ to denote the time slots during which each arriving task $z_i(t')$ is either processed or dropped in time slot $t$, as given by \vspace{3mm}

$\mathcal{F}_i^t =\bigg \{ t' \bigg|\; t' \leq t,\; \lambda_i(t')>0, \; (1 - x_i(t')) \; l_i^{\text{C}}(t') \;  $ 
\begin{equation}
	\hspace{17mm} +\,  x_i(t')\sum\limits_{j \in \mathcal{J}} \sum\limits_{n= t'}^{t}\mathbbm{1}(z_{i,j}^{\text{E}}(n)=z_i(t'))   l_{i,j}^{\text{E}}(n) = t \bigg\}.
	\label{28}  
	\nonumber
\end{equation}
Therefore, MD $i$ observes a set of QoEs $\{\boldsymbol{q}_i(t') \mid t' \in \mathcal{F}_i^t\}$ at the beginning of time slot $t+1$, where the set $\mathcal{F}_i^t$ for some $i \in \mathcal{I}$ can be empty. Subsequently, MD $i$ sends its experience $(\boldsymbol{s}_i(t), \boldsymbol{a}_i(t), \boldsymbol{q}_i(t), \boldsymbol{s}_i(t+1))$ to EN $j_i$ for each task $z_i(t')$ in $t' \in \mathcal{F}_i^t$.

\begin{algorithm}[tbp]
	\caption{\textcolor{black}{Training Process Algorithm at EN $j \in \mathcal{J}$}}\label{alg:cap2}
	\begin{algorithmic}[1]
		\renewcommand{\algorithmicrequire}{\textbf{\textcolor{black}{Input:}}} 
		\renewcommand{\algorithmicensure}{\textbf{\textcolor{black}{Output:}}}
		\Require \textcolor{black}{experience $(\hspace{-0.2mm}\boldsymbol{s}_i(\hspace{-0.2mm}t\hspace{-0.2mm}), \boldsymbol{a}_i(\hspace{-0.2mm}t\hspace{-0.2mm}), \boldsymbol{q}_i(\hspace{-0.2mm}t\hspace{-0.2mm}), \boldsymbol{s}_i(\hspace{-0.2mm}t\hspace{-0.3mm}+\hspace{-0.3mm}\hspace{-0.2mm}1\hspace{-0.2mm}\hspace{-0.2mm})\hspace{-0.2mm})$ from MD $i \in \mathcal{I}$}
		\Ensure \textcolor{black}{parameters vector $\theta_i^{\text{E}}$ }
		\State Initialize replay buffer $\mathcal{M}_i$ for each MD $i \in \mathcal{I}_j$;
		\State Initialize $\textit{Net}_i^{\text{E}}$ and $\textit{Net}_i^{\text{T}}$ with random parameters $\theta_i^{\text{E}}$ and $\theta_i^{\text{T}}$ respectively, for each MD $i \in \mathcal{I}_j$;
		\State Set Count := 0
		\While{True} \Comment{\textit{infinite loop}}
		\If{receive an \textit{UpdateRequest} from MD $i \in \mathcal{I}_j$}
		\State Send $\theta_i^{\text{E}}$ to MD $i \in \mathcal{I}$;
		\EndIf
		\If {an experience \hspace{-0.2mm} $(\hspace{-0.2mm}\boldsymbol{s}_i(\hspace{-0.2mm}t\hspace{-0.2mm}), \boldsymbol{a}_i(\hspace{-0.2mm}t\hspace{-0.2mm}), \boldsymbol{q}_i(\hspace{-0.2mm}t\hspace{-0.2mm}), \boldsymbol{s}_i(\hspace{-0.2mm}t\hspace{-0.3mm}+\hspace{-0.3mm}\hspace{-0.2mm}1\hspace{-0.2mm}\hspace{-0.2mm})\hspace{-0.2mm})$ is received \\ \;\;\;\; from MD $i \in \mathcal{I}_j$}
		\State Store $(\boldsymbol{s}_i(t'), \boldsymbol{a}_i(t'), \boldsymbol{q}_i(t'), \boldsymbol{s}_i(t'\hspace{-1mm}+1))$ in $\mathcal{M}_i$;
		\State Get a collection of experiences  $\mathcal{I}$ from $\mathcal{M}_i$; 
		\For{each experience $i \in \mathcal{I}$} 
		\State Get experience $(\boldsymbol{s}_i(\hspace{-0.2mm}n\hspace{-0.2mm}), \boldsymbol{a}_i(\hspace{-0.2mm}n\hspace{-0.2mm}), \boldsymbol{q}_i(\hspace{-0.2mm}n\hspace{-0.2mm}), \boldsymbol{s}_i(\hspace{-0.2mm}n\hspace{-0.5mm}+1\hspace{-0.2mm}))$; 
		\State Generate $\hat{Q}_{i,n}^{\text{T}}$ according to   \eqref{28};
		\EndFor
		\State Set vector  $\hat{\mathbf{Q}}_i^{\text{T}} := (\hat{Q}^{\text{T}}_{i,n})_{n \in \mathcal{N}}$;
		\State Update $\theta_i^{\text{E}}$ to minimize $L(\theta_i^{\text{E}}$, $\hat{\mathbf{Q}}_i^{\text{T}})$ in   \eqref{30};
		\State Count := Count + 1;
		\If {mod(Count, \textit{ReplaceThreshold}) = 0}
		\State $\theta_i^{\text{T}}$ := $\theta_i^{\text{E}}$;
		\EndIf
		\EndIf
		\EndWhile
		
	\end{algorithmic}
\end{algorithm}

\thispagestyle{empty}
\begin{textblock}{19.1}(1,1.4)
	\vspace{-5mm}
	\noindent \scriptsize \hspace{4mm}
	3126 \hfill IEEE TRANSACTIONS ON NETWORK SCIENCE AND ENGINEERING, VOL. 12, NO. 4, JULY/AUGUST 2025
\end{textblock}

\subsubsection{Training Process Algorithm at EN $j \in \mathcal{J}$}
Upon initializing the replay buffer $\mathcal{M}_i$ with the neural networks $\textit{Net}_i^{\text{E}}$ and $\textit{Net}_i^{\text{T}}$ for each MD $i \in \mathcal{I}_j$, EN $j \in \mathcal{J}$ waits for messages from the MDs in the set $\mathcal{I}_j$. When EN $j$ receives an $\textit{UpdateRequest}$ signal from an MD $i \in \mathcal{I}_j$, it responds by transmitting the updated parameter vector $\theta^{\text{E}}_i$, obtained from $\textit{Net}_i^{\text{E}}$, back to MD $i$. On the other side, if EN $j$ receives an experience $(\boldsymbol{s}_i(t), \boldsymbol{a}_i(t), \boldsymbol{q}_i(t), \boldsymbol{s}_i(t+1))$ from MD $i \in \mathcal{I}_j$, the EN stores this experience in the replay buffer $\mathcal{M}_i$ associated with that MD. 

The EN randomly selects a sample collection of experiences from the replay buffer, denoted as $\mathcal{N}$. For each experience $n \in \mathcal{N}$, it calculates the value of $	\hat{Q}^{\text{T}}_{i,n}$. This value represents the QoE in experience $n$ and includes a discounted Q-value of the action anticipated to be taken in the subsequent state of experience $n$, according to the network $\textit{Net}^\text{T}_i$, given by
\begin{alignat}{1}
	\hat{Q}_{i,n}^{\text{T}} = \boldsymbol{q}_i(n) + \gamma Q_i^{\text{T}}(\boldsymbol{s}_i(n+1)), \tilde{\boldsymbol{a}}_n; \theta_i^{\text{T}}),
	\label{29}  
\end{alignat}  
where $\tilde{\boldsymbol{a}}_n$ denotes the optimal action for the state $\boldsymbol{s}_i(n+1)$ based on its highest Q-value under $\textit{Net}_i^{\text{E}}$, as given by:
\begin{alignat}{1}
	\tilde{\boldsymbol{a}}_n = \text{arg} \; \max_{\boldsymbol{a} \in \mathcal{A}} \; Q_i^{\text{E}}(\boldsymbol{s}_i(n+1), \boldsymbol{a}; \theta_i^{\text{E}}).
	\label{30}  
\end{alignat} 
In particular, regarding experience $n$, the target-Q value $\hat{Q}_{i,n}^{\text{T}}$ represents the long-term QoE for action $\boldsymbol{a}_i(n)$ under state $\boldsymbol{s}_i(n)$. This value corresponds to the QoE observed in experience $n$, as well as the approximate expected upcoming QoE. 
\textcolor{black}{Based on the set $\mathcal{N}$, the EN computes the vector $\hat{\mathbf{Q}}_i^{\text{T}} = (\hat{Q}^{\text{T}}_{i,n})_{n \in \mathcal{N}}$ and trains the MD's neural network (Lines 11-21 of Algorithm~\ref{alg:cap2}) to update the parameter vector $\theta^{\text{E}}_i$ in $\textit{Net}_i^{\text{E}}$ for the next MD's \textit{UpdateRequest}.} The key idea of updating $\textit{Net}_i^{\text{E}}$ is to minimize the disparity in Q-values between $\textit{Net}_i^{\text{E}}$ and $\textit{Net}_i^{\text{T}}$, as indicated by the following loss function:
\begin{alignat}{1}
     L\Big(\theta_i^{\text{E}},\hat{\mathbf{Q}}_i^{\text{T}}\Big) = {1 \over |\mathcal{N}| } \sum\limits_{n \in \mathcal{N}} \bigg(Q_i^{\text{E}}(\boldsymbol{s}_i(n), \boldsymbol{a}_i(n); \theta_i^{\text{E}} ) - \hat{Q}^{\text{T}}_{i,n}  \bigg)^2.
	\label{31}  
\end{alignat}  
In every \textit{ReplaceThreshold} iterations, the update of $\textit{Net}_i^{\text{T}}$ will involve duplicating the parameters from $\textit{Net}_i^{\text{E}}$ ($\theta_i^{T} = \theta_i^{E}$). The objective is to consistently update the network parameter $\theta_i^{T}$ in $\textit{Net}_i^{\text{T}}$, which enhances the approximation of the long-term QoE when computing the target Q-values in \eqref{28}.\\

\subsubsection{Computational Complexity}

The computational complexity of the QECO algorithm is determined by the number of experiences required to discover the optimal offloading policy. Each experience involves backpropagation for training, which has a computational complexity of $\mathcal{O}(C)$, where $C$ represents the number of multiplication operations in the neural network. During each training round triggered by the arrival of a new task, a sample collection of experiences of size $|\mathcal{N}|$ is utilized from the replay buffer. Since the training process encompasses $N^{\text{ep}}$ episodes and there are $K$ expected tasks in each episode, the computational complexity of the proposed algorithm is $\mathcal{O}(N^{\text{ep}}K|\mathcal{N}|C$), which is polynomial. Given the integration of neural networks for function approximation, the convergence guarantee of the DRL algorithm remains an open problem. In this work, we will empirically evaluate the convergence of the proposed algorithm in Section \ref{section:2}.


\section{Performance Evaluation}\label{section:VI}
In this section, we first present the simulation setup and training configuration. We then evaluate the performance of QECO in comparison to three baseline schemes in addition to the existing state-of-the-art works~\Cite{yang2018distributed},~\Cite{qiu2020distributed}. We further analyze the convergence of our proposed algorithm. 

\subsection{Simulation Setup}
We consider a MEC environment with 50 MDs and 5 ENs, similar to \Cite{9253665}. We also follow the model presented in \Cite{zhou2021deep} to determine the energy consumption. All the parameters are given in Table~\ref{table}. \textcolor{black}{Since there is no real dataset due to the challenges of capturing representative samples over extended periods, we use a simulated MEC system that enables DRL agents to continuously gather experience and improve their performance based on system feedback (e.g., QoE).} To train the MDs' neural networks, we adopt a scenario comprising 1000 episodes. Each episode contains 100 time slots, each of length 0.1 second. The QECO algorithm incorporates real-time experience into its training process to continuously enhance the offloading strategy. Specifically, we employ a batch size of 16, maintain a fixed learning rate of 0.001, and set the discount factor $\gamma$ to 0.9. The probability of random exploration gradually decreases from an initial value 1, progressively approaching 0.01, all of which is facilitated by an RMSProp optimizer. \textcolor{black}{The algorithm's source code is available at \Cite{QECO}}.

\thispagestyle{empty}
\begin{textblock}{19.1}(1,1.4)
	\vspace{-5mm}
	\noindent \scriptsize \hspace{4mm}
	RAHMATY et al.: QECO: A QOE-ORIENTED COMPUTATION OFFLOADING ALGORITHM BASED ON DEEP REINFORCEMENT LEARNING \hfill 3127
\end{textblock}

	\begin{table}
	\renewcommand{\arraystretch}{1.1}\centering
	\captionsetup{name=TABLE}
	\caption{Simulation Parameters}
	\scalebox{0.93}{%
		\begin{tabular}{ l|l } 
			\toprule
			\textbf{Parameter} & \textbf{Value}  \\ 
			\midrule
			Computation capacity of MD $f_i$  & 2.6 GHz \\ 
			Computation capacity of EN $f_j^{\text{E}}$ & 42.8 GHz  \\ 
			Transmission capacity of MD $r_{i,j}(t)$ & 14 Mbps  \\ 
			Task arrival rate & 150 Task/sec  \\ 
			Size of task $\lambda_i(t)$  & $\{\text{1.0, 1.1, \ldots , 7.0}\}$ Mbits \\ 
			Required CPU cycles of task $\rho_i(t)$  & $\{\text{0.197, 0.297, 0.397}\}$ $\times10^3$ \\ 
			Deadline of task $ \Delta_i$  & 10 time slots (1 Sec)\\ 
			\textcolor{black}{Percentage of MD $i$'s battery level} $\phi_i(t)$  & $\{\text{25, 50, 75}\}$\\ 
			Computation power of EN $p_j^{\text{E}}$ & 5 W  \\ 
			Transmission power of MD $p_i^{\text{T}}$ & 2.3 W  \\ 
			Standby power of MD $p_i^{\text{I}}$& 0.1 W  \vspace{1mm}\\ 
			\toprule
	\end{tabular}}\vspace{-2mm}
	\label{table}
\end{table}


\hspace{-2mm}We use the following methods as benchmarks.

\begin{enumerate}

\item \textit{Local Computing (LC):} The MDs execute all of their computation tasks using their own computing capacity.

\item \textit{Full Offloading (FO):} Each MD dispatches all of its computation tasks while choosing the offloading target randomly. 

\item \textit{Random Decision (RD):} In this approach, when an MD receives a new task, it randomly makes the offloading decisions and selects the offloading target if it decides to dispatch the task. 

\item \textit{PGOA~\Cite{yang2018distributed}: }
This existing method is a distributed optimization algorithm designed for delay-sensitive tasks in an environment where MDs interact strategically with multiple ENs. 

\item \textit{DCDRL~\Cite{qiu2020distributed}:} This method is based on the AC framework \Cite{NIPS1999_6449f44a}, which underpins many state-of-the-art DRL algorithms, such as DDPG \Cite{lillicrap2015continuous}, and PPO \Cite{schulman2017proximal}. DCDRL is designed for distributed computation offloading in a queuing-based MEC environment. 
\end{enumerate}





\begin{figure}[tbp]
	\captionsetup{name=Fig.}
	\begin{minipage}[b]{0.5\linewidth}
		\centering
		\includegraphics[width=\textwidth]{ 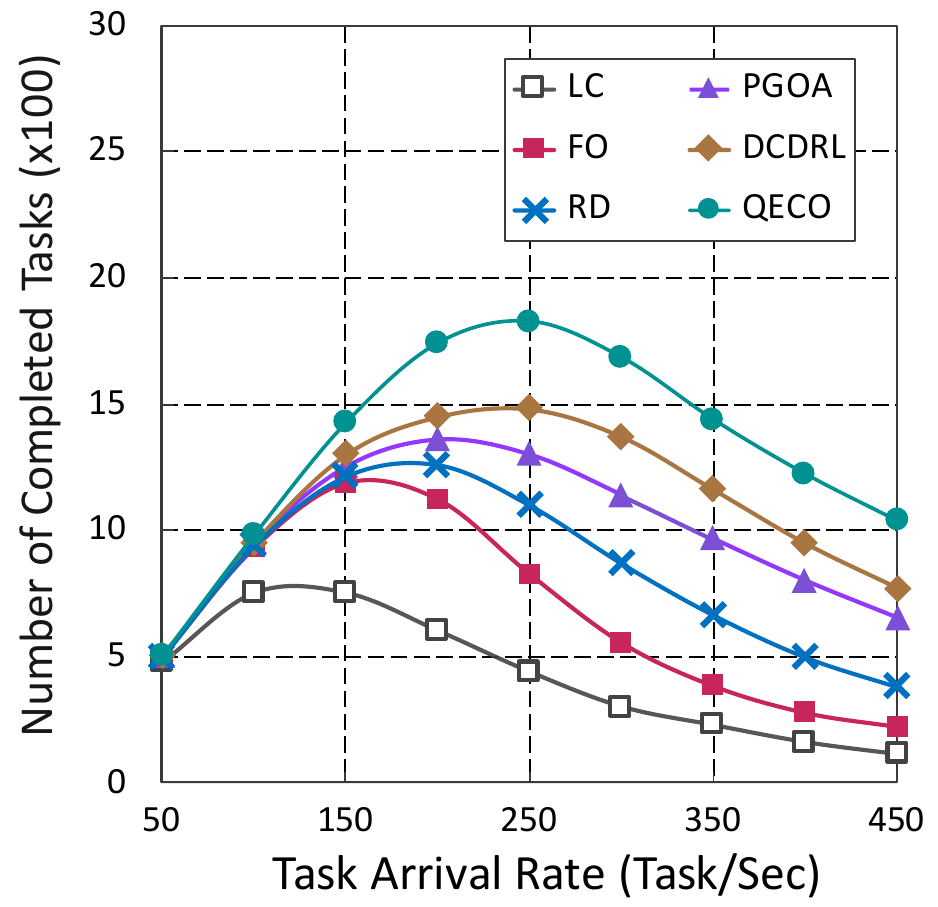} 
		\textcolor{white}{i}\hspace{0.6cm}(a)
	\end{minipage}
	\hspace{-0.2cm}
	\begin{minipage}[b]{0.5\linewidth}
		\centering
		\includegraphics[width=\textwidth]{ 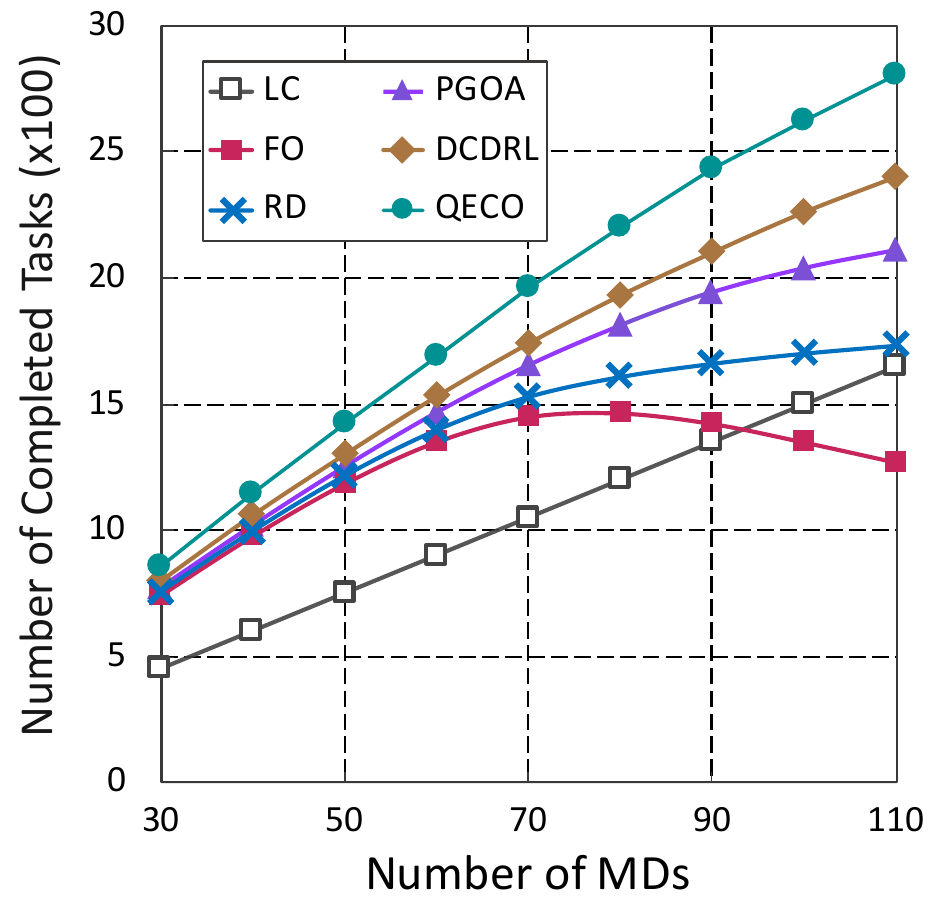}
		\textcolor{white}{i}\hspace{0.6cm}(b)
	\end{minipage}\vspace{-1mm}
	\vspace{-0.5cm}
	\caption{\textcolor{black}{The number of completed tasks under different computation workloads: (a) task arrival rate; (b) the number of MDs.}}
	\label{chart1}
\end{figure}

\begin{figure}[tbp]
	\captionsetup{name=Fig.}
	\begin{minipage}[b]{0.50\linewidth}
		\centering
		\includegraphics[width=\textwidth]{ 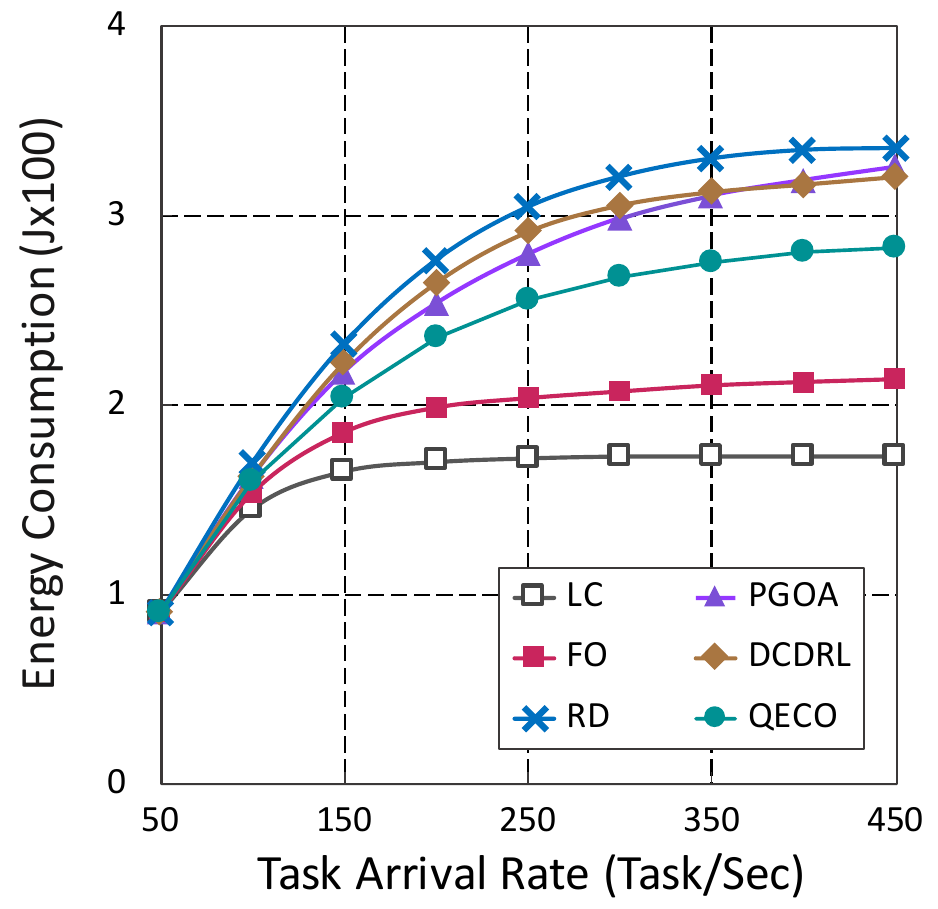} 		
		\textcolor{white}{i}\hspace{0.6cm}(a)
	\end{minipage}
	\hspace{-0.2cm}
	\begin{minipage}[b]{0.50\linewidth}
		\centering
		\includegraphics[width=\textwidth]{ 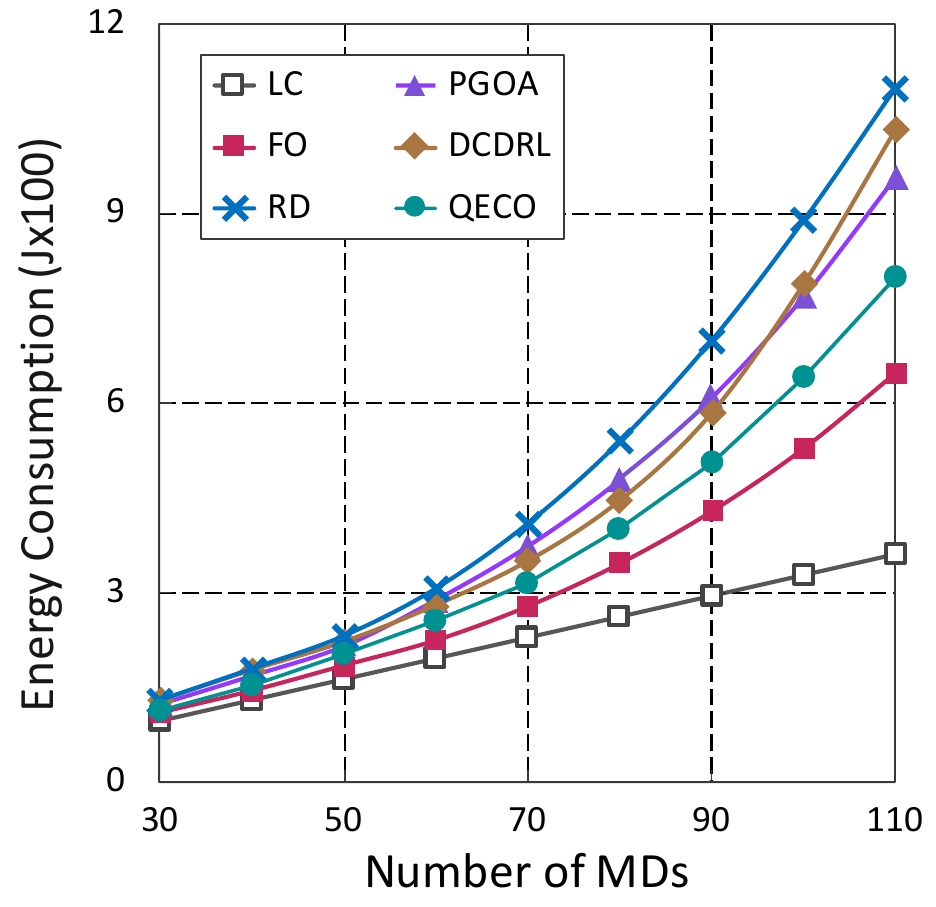}
		\textcolor{white}{i}\hspace{0.6cm}(b)
	\end{minipage}\vspace{-1mm}
	\vspace{-0.5cm}
	\caption{\textcolor{black}{The overall energy consumption under different computation workloads: (a) task arrival rate; (b) the number of MDs.}}\vspace*{-3mm}
	\label{chart2}
\end{figure}

\vspace{-3mm}
\subsection{Performance Comparison and Convergence}
\label{section:2}

We first evaluate the number of completed tasks when comparing our proposed QECO algorithm with the other four schemes. As illustrated in Fig.~\ref{chart1}(a), the QECO algorithm consistently outperforms the benchmark methods when we vary the task arrival rate. At a lower task arrival rate (i.e., 50), most of the methods demonstrate similar proficiency in completing tasks. However, as the task arrival rate increases, the efficiency of QECO becomes more evident. Specifically, when the task arrival rate increases to 250, our algorithm can increase the number of completed tasks by 72.1\%, 46.8\%, and 28.6\% compared to RD, PGOA, and DCDRL, respectively.
Similarly, in Fig.~\ref{chart1}(b), as the number of MDs increases, QECO shows significant improvements in the number of completed tasks compared to other methods, especially when faced with a large number of MDs. When there are 110 MDs, our proposed algorithm can effectively increase the number of completed tasks by at least 22.8\% comparing to other methods. This achievement is attributed to the QECO's ability to effectively handle unknown workloads and prevent congestion at the ENs.

Figs.~\ref{chart2}(a) and~\ref{chart2}(b) illustrate the overall energy consumption for different values of task arrival rate and the number of MDs, respectively. At the lower task arrival rate, the total energy consumption of all methods is close to each other. The total energy consumption increases when we have a higher task arrival rate.  
As can be observed from Fig.~\ref{chart2}(a), at task arrival rate 450, QECO effectively reduces overall energy consumption by 18.6\%, 15.5\%, and 13.9\% compared to RD, PGOA, and DCDRL, respectively, as it takes into account the \textcolor{black}{energy level} of the MD in its decision-making process. However, it consumes more energy compared to LC and FO because they do not utilize all computing resources. In particular, LC only uses the MD's computational resources, while FO utilizes the allocated EN computing resources. 
In Fig.~\ref{chart2}(b), an increasing trend in overall energy consumption is observed as the number of MDs increases since the number of resources available in the system increases, which leads to higher energy consumption. The QECO algorithm consistently outperforms other methods in overall energy consumption, especially when there are a large number of MDs. Specifically, QECO demonstrates a 27.4\%, 16.7\%, and 23.5\% reduction in overall energy consumption compared to RD, PGOA, and DCDRL, respectively, when the number of MDs increases to 110.

\begin{figure}[t]
	\captionsetup{name=Fig.}
	\begin{minipage}[b]{0.50\linewidth}
		\centering
		\includegraphics[width=\textwidth]{ 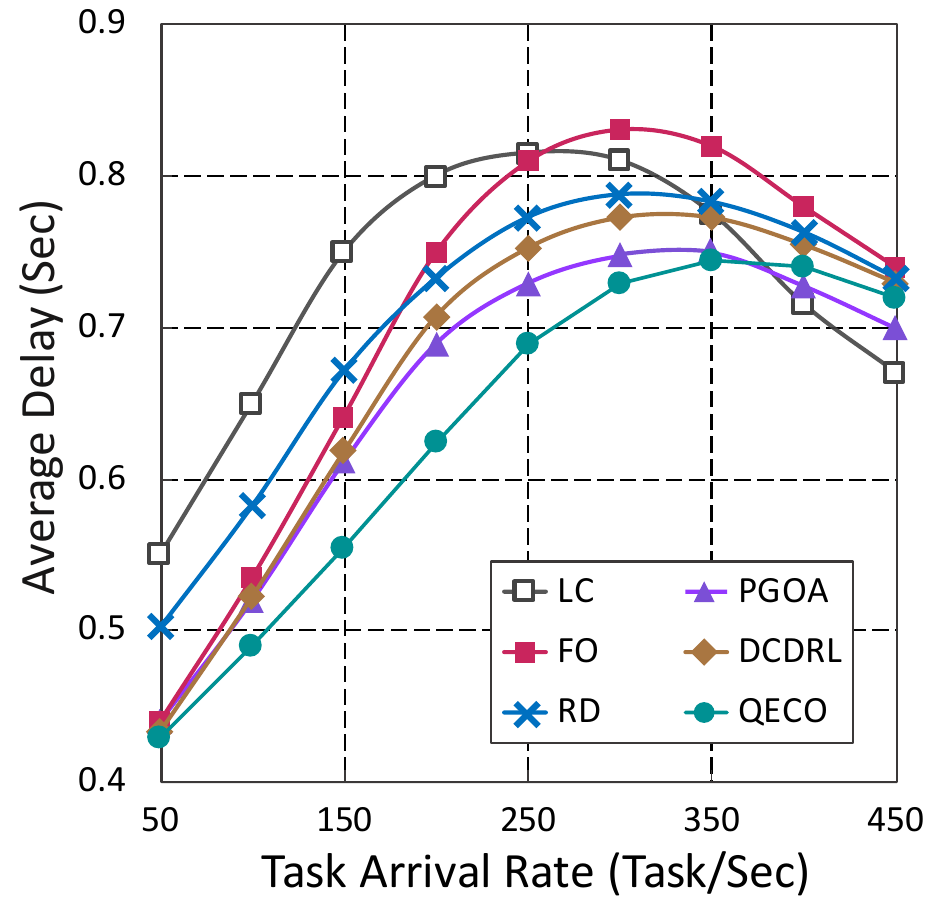} 		
		\textcolor{white}{i}\hspace{0.6cm}(a)
	\end{minipage}
	\hspace{-0.2cm}
	\begin{minipage}[b]{0.50\linewidth}
		\centering
		\includegraphics[width=\textwidth]{ 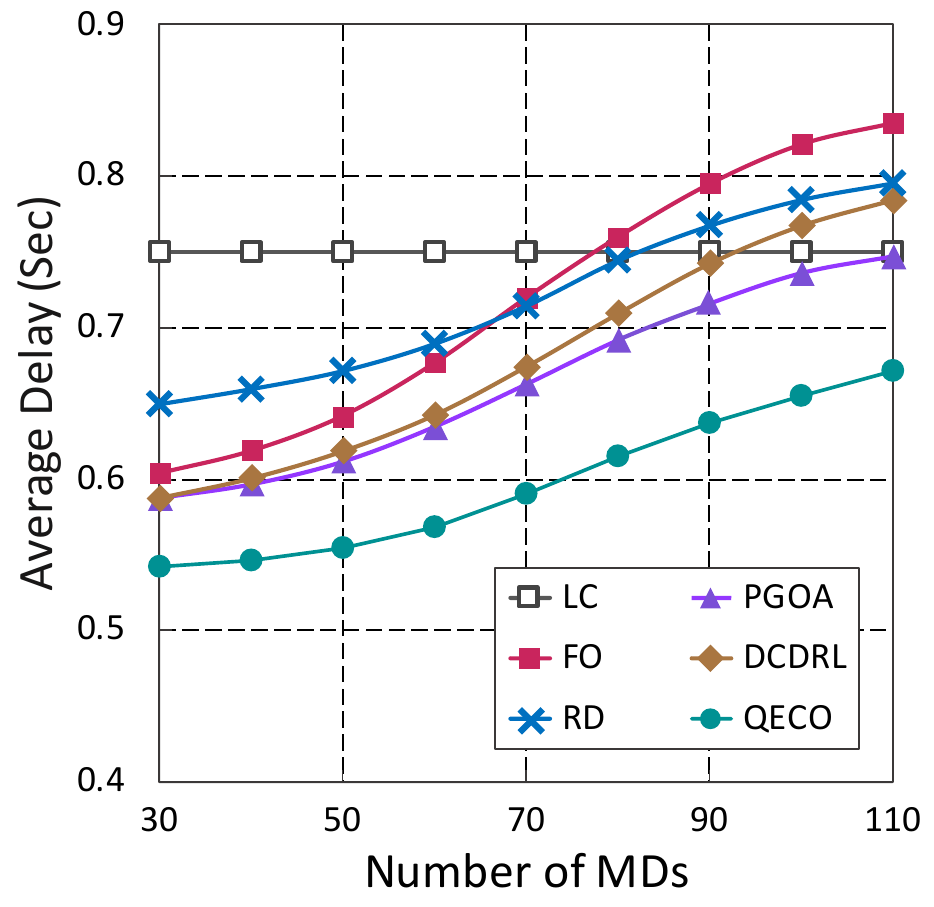}
		\textcolor{white}{i}\hspace{0.6cm}(b)
	\end{minipage}\vspace{-1mm}
	\vspace{-0.5cm}
	\caption{\textcolor{black}{The average delay under different computation workloads: (a) task arrival rate; (b) the number of MDs.}}
		\vspace*{-3mm}
	\label{chart3}
\end{figure}

As shown in Fig.~\ref{chart3}(a), the QECO algorithm maintains a lower average delay compared to other methods as the task arrival rate increases from 50 to 350. Specifically, when the task arrival rate is 200, it reduces the average delay by at least 12.4\% compared to other methods. However, for task arrival rates exceeding 350, QECO may experience a higher average delay compared to some of the other methods. This can be attributed to the fact that the other algorithms drop more tasks while our proposed algorithm is capable of completing a higher number of tasks, potentially leading to an increase in average delay. In Fig.~\ref{chart3}(b), as the number of MDs increases, we observe a rising trend in the average delay. It can be inferred that an increase in computational load in the system can lead to higher queuing delays and computations at ENs. Considering the QECO's ability to schedule workloads, when the number of MDs increases from 30 to 110, it consistently maintains a lower average delay which is at least 8.3\% less than the other methods.

\begin{figure}[tbp]
 	\captionsetup{name=Fig.}
 	\begin{minipage}[b]{0.50\linewidth}
 		\centering
 		\includegraphics[width=\textwidth]{ 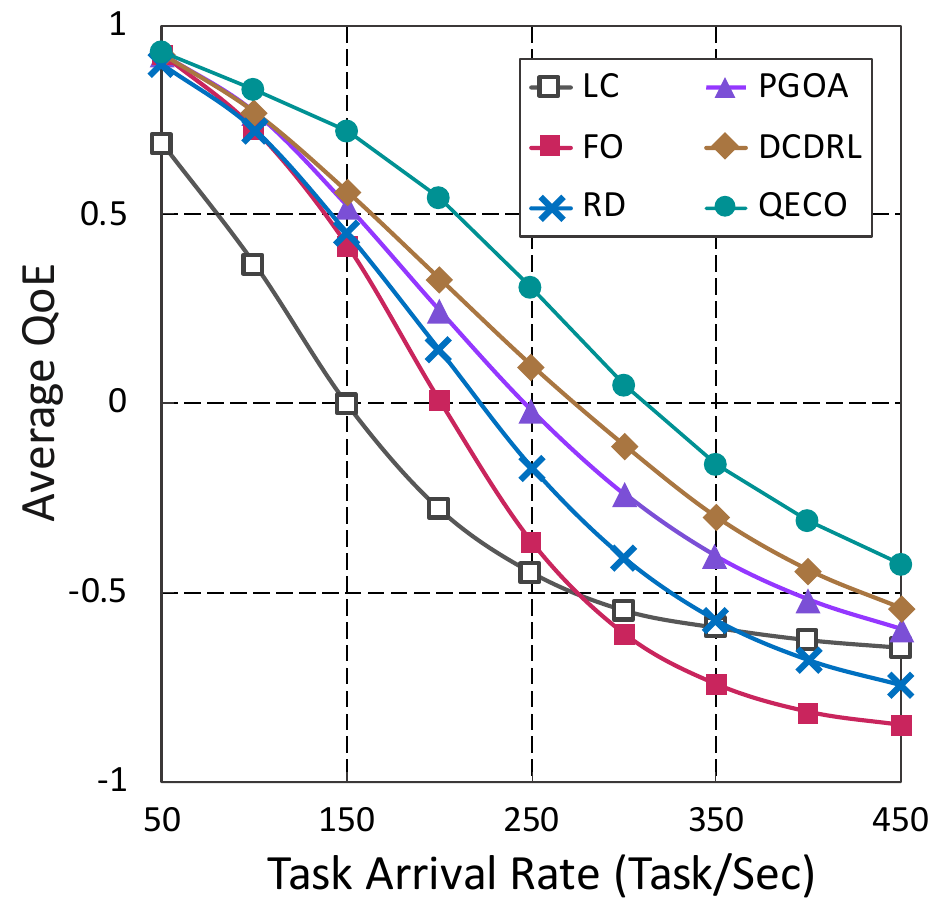} 		
 		\textcolor{white}{i}\hspace{0.6cm}(a)
 	\end{minipage}
 	\hspace{-0.2cm}
 	\begin{minipage}[b]{0.50\linewidth}
 		\centering
 		\includegraphics[width=\textwidth]{ 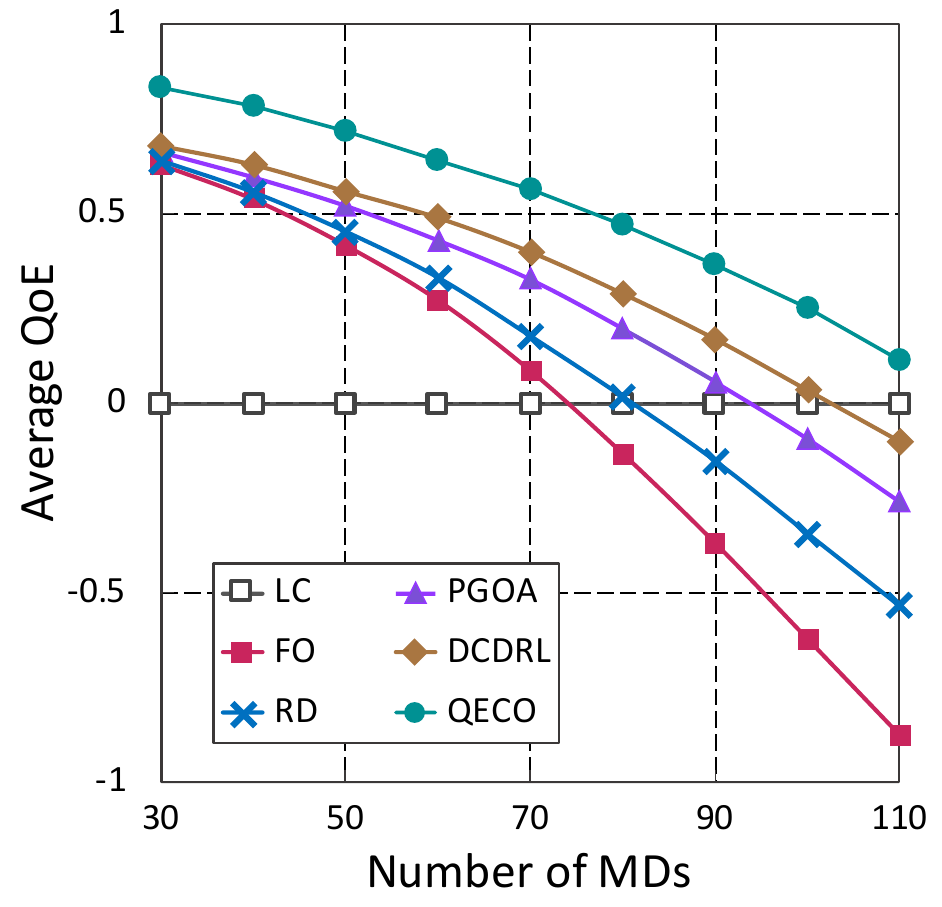}
 		\textcolor{white}{i}\hspace{0.6cm}(b)
 	\end{minipage}\vspace{-1mm}
 	\vspace{-0.5cm}
 	\caption{\textcolor{black}{The average QoE under different computation workloads: (a) task arrival rate; (b) the number of MDs.}}
 	\label{chart4}
 \end{figure}

We further investigate the overall improvement achieved by the QECO algorithm in comparison to other methods in terms of the average QoE. This metric signifies the advantages MDs obtain by utilizing different algorithms. \textcolor{black}{Fig.~\ref{chart4}(a) shows the average QoE for different values of the task arrival rate. This figure highlights the superiority of the QECO algorithm in providing MDs with an enhanced experience. At lower task arrival rates (i.e., 50-150), QECO maintains an average QoE of at least 0.72, while the other methods experience a steeper decline, with average QoE dropping to 0--0.56. Specifically, when the task arrival rate is 200, QECO improves the average QoE by at least 65.7\% compared to other methods. As task arrival rates increase to 300, the average QoE significantly decreases for all methods due to increased competition for resources in the MEC system. However, QECO still maintains a positive average QoE, while other methods fall to negative values. At higher task arrival rates (i.e., 350–450), QECO achieves 42.6\%, 28.5\%, and 22.6\% improvement in QoE compared to RD, PGOA, and DCDRL, respectively.}

 Fig.~\ref{chart4}(b) illustrates the average QoE when we increase the number of MDs. The EN's workload grows when there are a larger number of MDs, leading to a reduction in the average QoE of all methods except LC. However, QECO effectively manages the uncertain load at the ENs. \textcolor{black}{When the number of MDs increases from 30 to 110, QECO consistently maintains at least a 24.8\% higher QoE compared to the other methods. Specifically, at a moderate number of 70 MDs, QECO achieves an average QoE of 0.57, showing a 70.3\% and 44.7\% improvement compared to PGOA and DCDRL, respectively.} It is worth noting that although improvements in each of the QoE factors can contribute to enhancing system performance, it is essential to consider the user's demands in each time slot. Therefore, the key difference between QECO and other methods is that it prioritizes users' demands, enabling it to strike an appropriate balance among them, ultimately leading to a higher QoE for~MDs.
 
 \thispagestyle{empty}
 \begin{textblock}{19.1}(1,1.4)
 	\vspace{-5mm}
 	\noindent \scriptsize \hspace{4mm}
 	3128 \hfill IEEE TRANSACTIONS ON NETWORK SCIENCE AND ENGINEERING, VOL. 12, NO. 4, JULY/AUGUST 2025
 \end{textblock}
 
  \vspace{-1mm}

\begin{figure}
	\captionsetup{name=Fig.}
	\centering
	\includegraphics[width=0.88\linewidth]{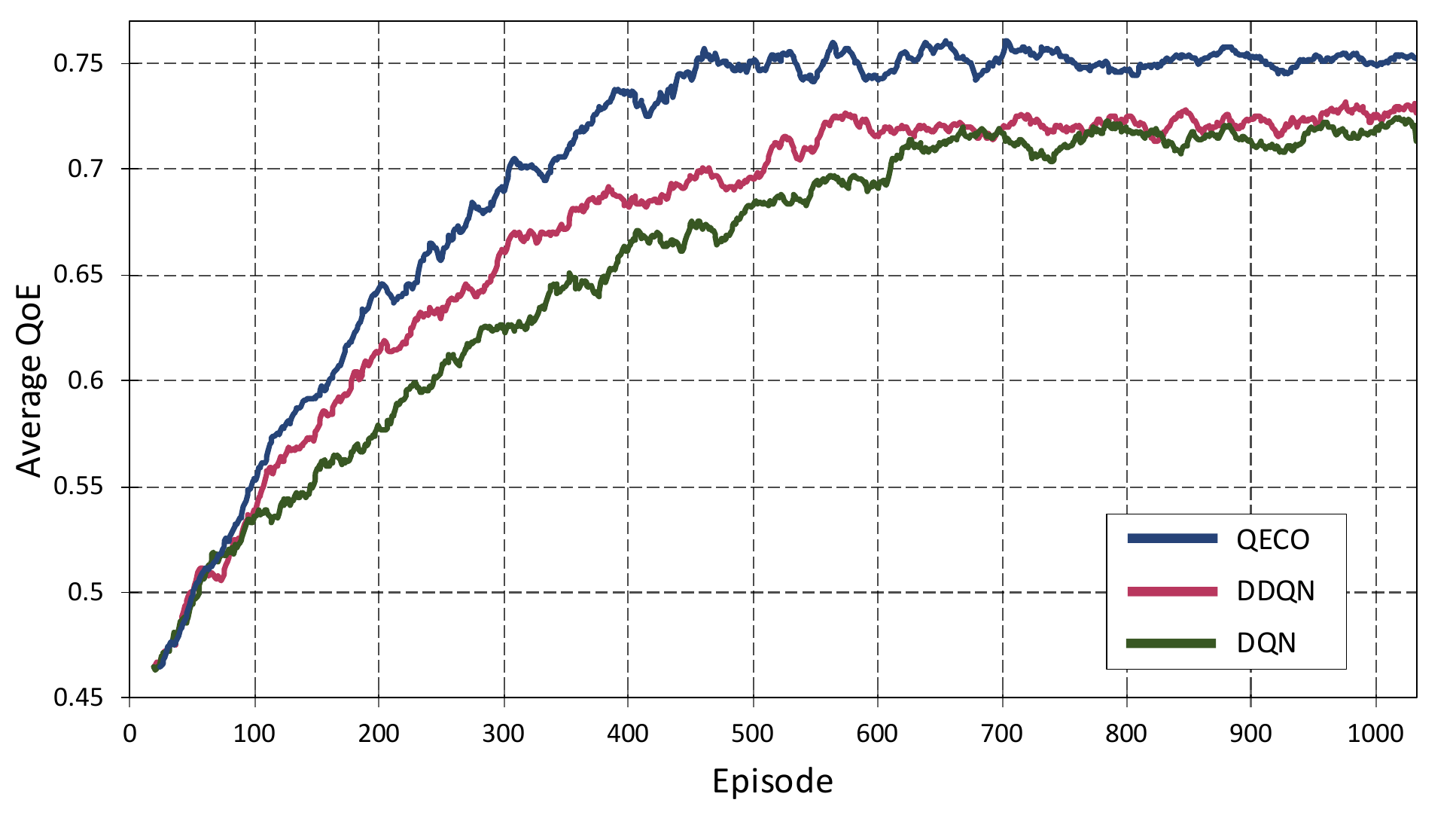}
	\vspace*{-3mm}
	\caption{\textcolor{black}{The convergence of the average QoE across episodes under different DQN-based methods.
	}}
	\vspace*{-3mm}
	\label{chart01}
\end{figure}

\begin{figure}
	\captionsetup{name=Fig.}
	\begin{minipage}[b]{0.5\linewidth}
		\centering
		\includegraphics[width=\textwidth]{ 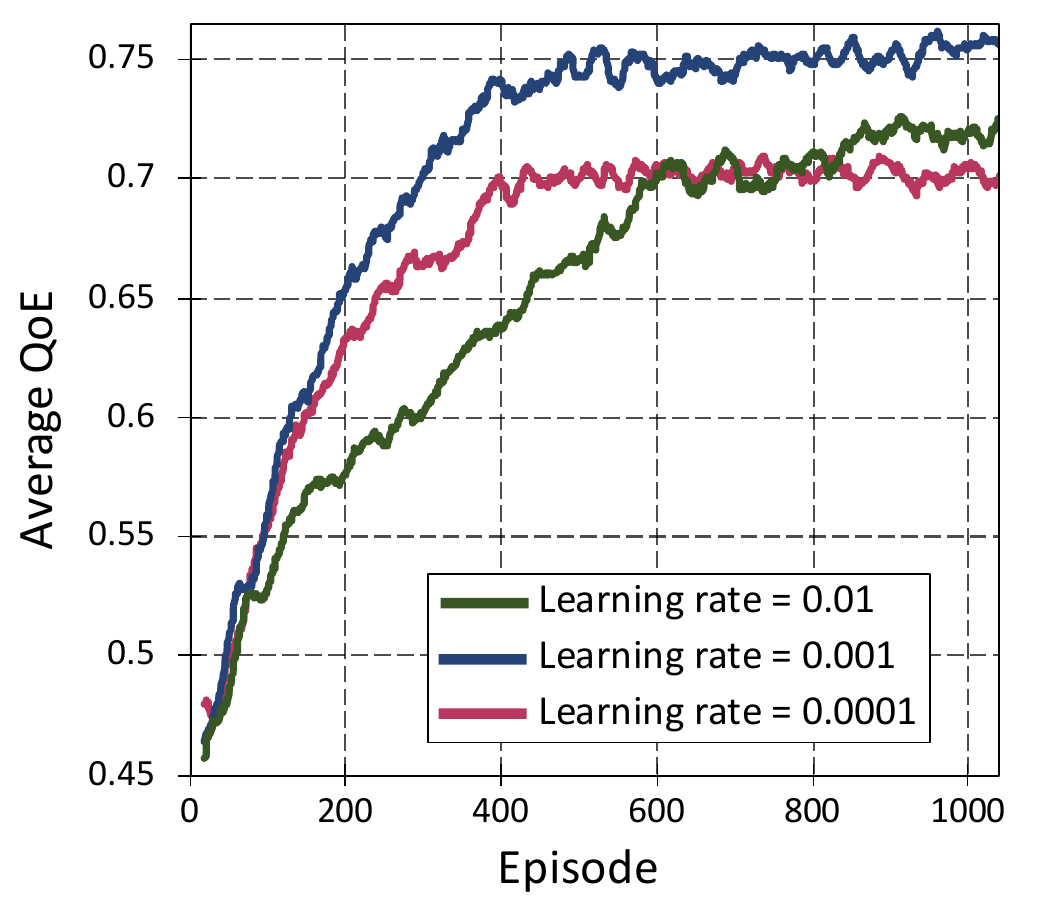} 
		\textcolor{white}{i}\hspace{0.6cm}(a)
	\end{minipage}
	\hspace{-0.26cm}
	\begin{minipage}[b]{0.5\linewidth}
		\centering
		\includegraphics[width=\textwidth]{ 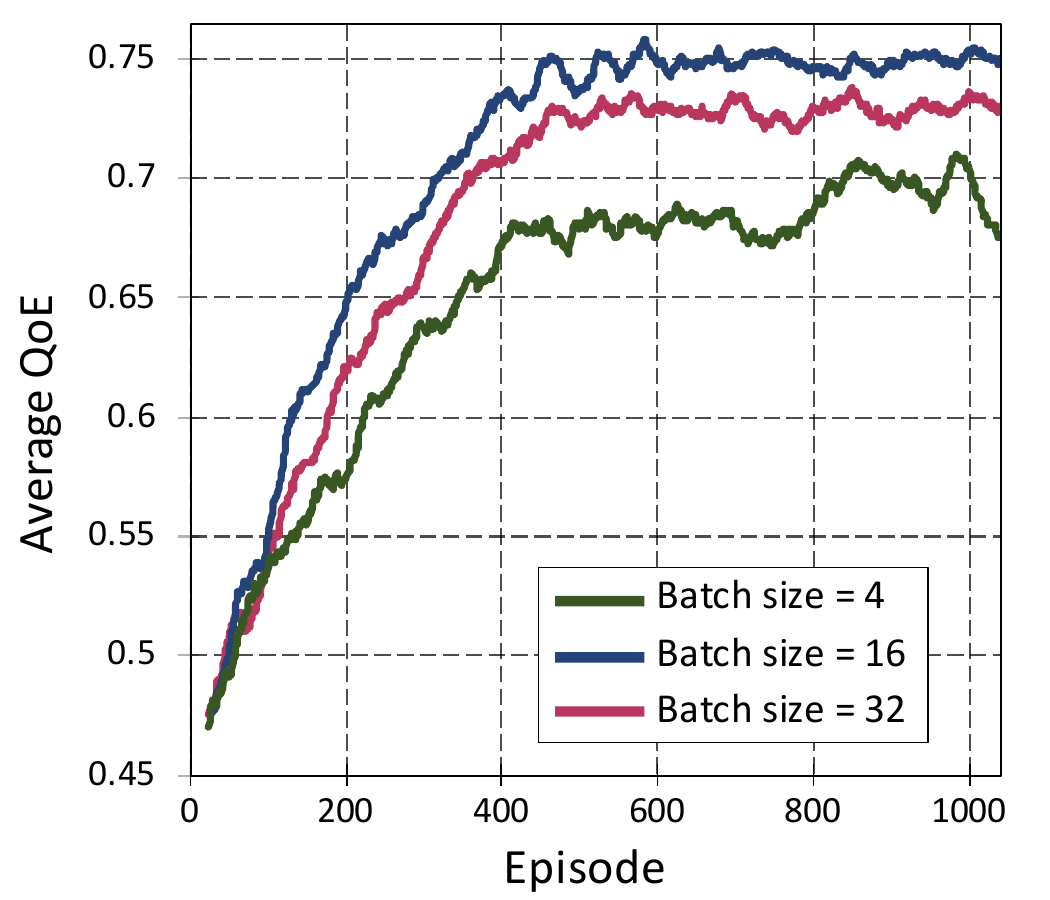}
		\textcolor{white}{i}\hspace{0.6cm}(b)
	\end{minipage}\vspace{-1mm}
	\vspace{-0.5cm}
	\caption{\textcolor{black}{The convergence of the average QoE across episodes under different hyper-parameters: (a) Learning rate; (b) Batch size.}}
		\vspace*{-2mm}
	\label{chart02}
\end{figure}

\textcolor{black}{	
We finally delve into the investigation of the convergence performance of the QECO algorithm in Fig.~\ref{chart01} and Fig.~\ref{chart02}. To validate the effectiveness of the QECO algorithm, we assess its convergence rate compared to the vanilla DQN and DDQN configurations \Cite{van2016deep}, measured by the average QoE across episodes. As shown in Fig.~\ref{chart01}, the MD's network progressively learns efficient policies over the episodes, ultimately stabilizing as it approaches convergence. Specifically, DQN, DDQN, and our proposed QECO algorithm converge after approximately 650, 550, and 450 iterations, respectively. 
The QECO algorithm demonstrates the faster convergence while achieving a higher average QoE than the other methods. This underscores the beneficial impact of workload prediction by the LSTM network and highlights its effectiveness in efficiently utilizing the computing resources of MDs and ENs.
}

Furthermore, we explore the impact of two main hyper-parameters on the convergence speed and the converged result of the proposed algorithm. Fig.~\ref{chart02}(a) illustrates the convergence of the proposed algorithm under different learning rates, where the learning rate regulates the step size per iteration towards minimizing the loss function. The QECO algorithm achieves an average QoE of 0.75 when the learning rate is 0.001, indicating relatively rapid convergence. However, with smaller learning rates (e.g., 0.0001) or larger values (e.g., 0.01),  a slower convergence is observed. Fig.~\ref{chart02}(b) shows the convergence of the proposed algorithm under different batch sizes, which refer to the number of sampled experiences in each training round. An improvement in convergence performance is observed as the batch size increases from 4 to 16. However, further increasing the batch size from 16 to 32 does not notably enhance the converged QoE or convergence speed. Hence, a batch size of 16 may be more appropriate for training processes.

\thispagestyle{empty}
\begin{textblock}{19.1}(1,1.4)
	\vspace{-5mm}
	\noindent \scriptsize \hspace{4mm}
	RAHMATY et al.: QECO: A QOE-ORIENTED COMPUTATION OFFLOADING ALGORITHM BASED ON DEEP REINFORCEMENT LEARNING \hfill 3129
\end{textblock}

\section{Conclusion}

\label{section:VII}
In this paper, we focused on addressing the challenge of offloading in MEC systems, where strict task processing deadlines and energy constraints adversely impact system performance. We formulated an optimization problem that aims to maximize the QoE of each MD individually, while QoE reflects the energy consumption and task completion delay. To address the dynamic and uncertain mobile environment, we proposed a QoE-oriented DRL-based computation offloading algorithm called QECO. Our proposed algorithm empowers MDs to make offloading decisions without relying on knowledge about task models or other MDs' offloading decisions. The QECO algorithm not only adapts to the uncertain dynamics of load levels at ENs, but also effectively manages the ever-changing system environment. Through extensive simulations, we showed that QECO outperforms several established benchmark techniques, while demonstrating a rapid training convergence. Specifically, QECO increases the average user's QoE by 29.8\%\,--\,37.1\% compared to several existing algorithms. This advantage can lead to improvements in key performance metrics, including task completion rate, task delay, and energy consumption, under different system conditions and varying user demands.

There are multiple directions for future work. A complementary approach involves extending the task model by considering interdependencies among tasks. This can be achieved by incorporating a task call graph representation. Furthermore, in order to accelerate the learning of optimal offloading policies, it will be beneficial to take advantages of federated learning techniques in the training process. This will allow MDs to collectively contribute to improving the offloading model and enable continuous learning when new MDs join the network.

\bibliographystyle{IEEEtranN} 
\bibliography{paper} 
\begin{IEEEbiography}[{\includegraphics[width=1in,height=1.25in,clip,keepaspectratio]{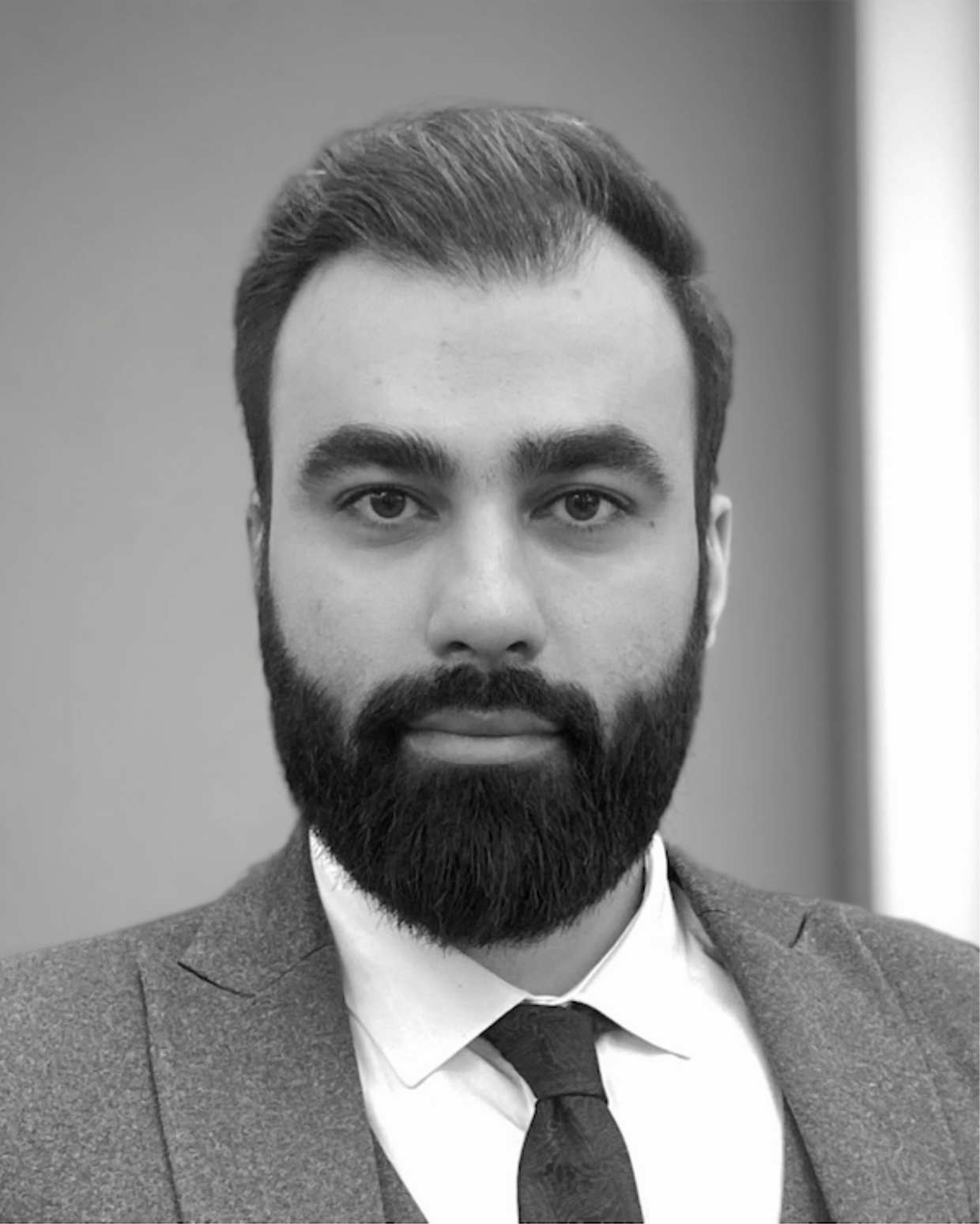}}]{Iman Rahmaty}
	received the M.Sc. degree in computer science and engineering from Sharif University of Technology, Tehran, Iran, in 2022, he was a Research Assistant with the Performance and Dependability Laboratory. He was with the Department of Electrical Engineering, Sharif University of Technology, in 2023, where he is currently a Research Engineer in the EdgeAI laboratory. His research interests include applying AI/ML-driven solutions, including reinforcement learning, federated learning, and meta-learning to address the challenges and complexities of distributed systems, Internet of Things, mobile edge computing, and wireless networks.
\end{IEEEbiography}	
\begin{IEEEbiography}
	[{\includegraphics[width=1in,height=1.25in,clip,keepaspectratio]{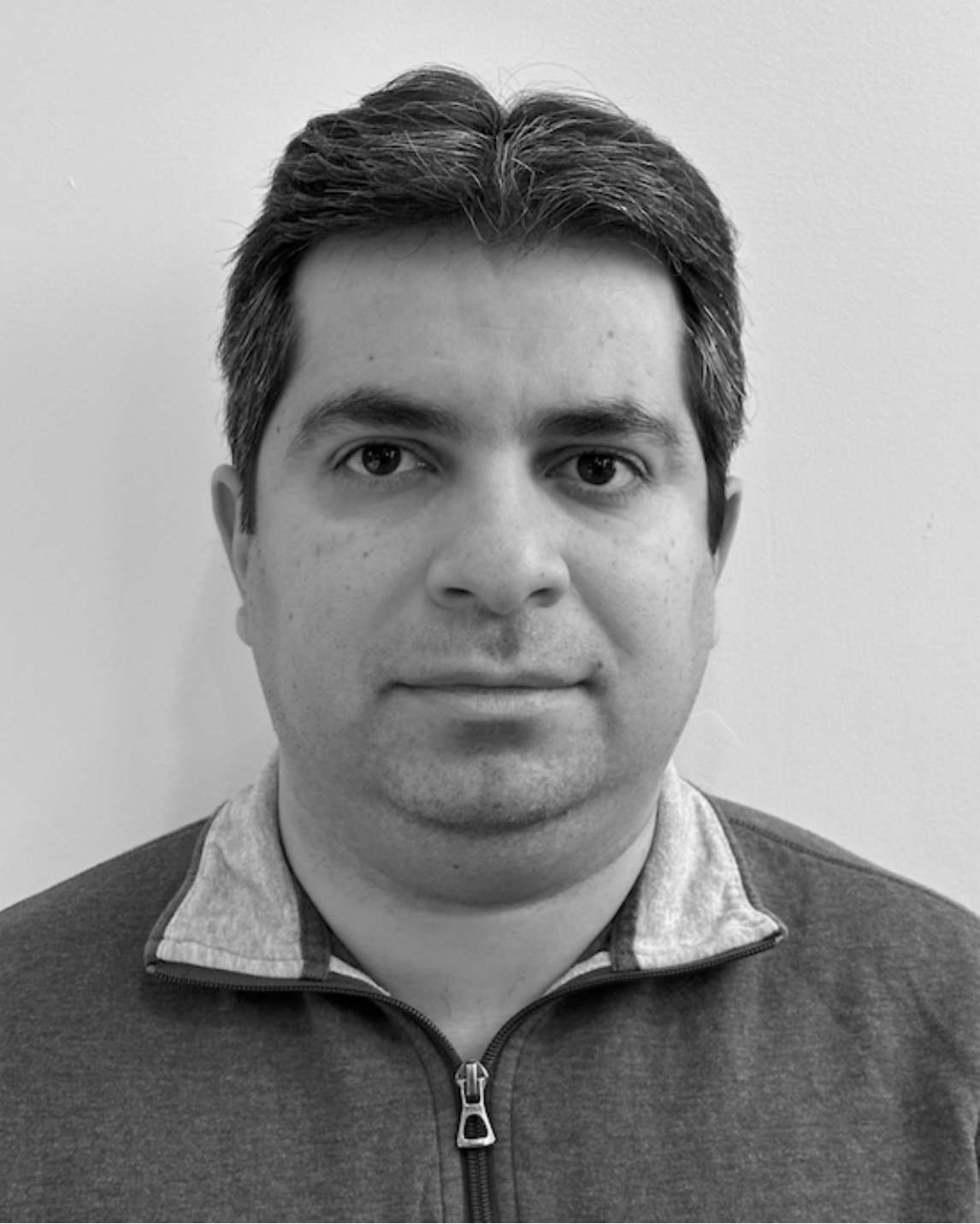}}]{Hamed Shah-Mansouri} (Member, IEEE) received the B.Sc., M.Sc., and Ph.D. degrees in electrical engineering from the Sharif University of Technology, Tehran, Iran, in 2005, 2007, and 2012, respectively. He was a Postdoctoral Research and a Teaching Fellow with the University of British Columbia, Vancouver, BC, Canada. He was with the Department of Electrical Engineering, Sharif University of Technology, in 2019 where he is currently an Assistant Professor. He has served as the publication co-chair for the {\it{IEEE Canadian Conference on Electrical and Computer Engineering}} 2016 and {\it{IEEE International Conference on Systems, Man and Cybernetics}} 2020 and as the technical program committee (TPC) member for several conferences including the {\it{IEEE Global Communications Conference (GLOBECOM)} 2015} and the {\it{IEEE Vehicular Technology Conference (VTC--Fall) 2016, 2017, and 2018}}. His research interests include the application of optimization and machine learning in computer networks, and Internet of Things. 
\end{IEEEbiography}	
\begin{IEEEbiography}
	[{\includegraphics[width=1in,height=1.25in,clip,keepaspectratio]{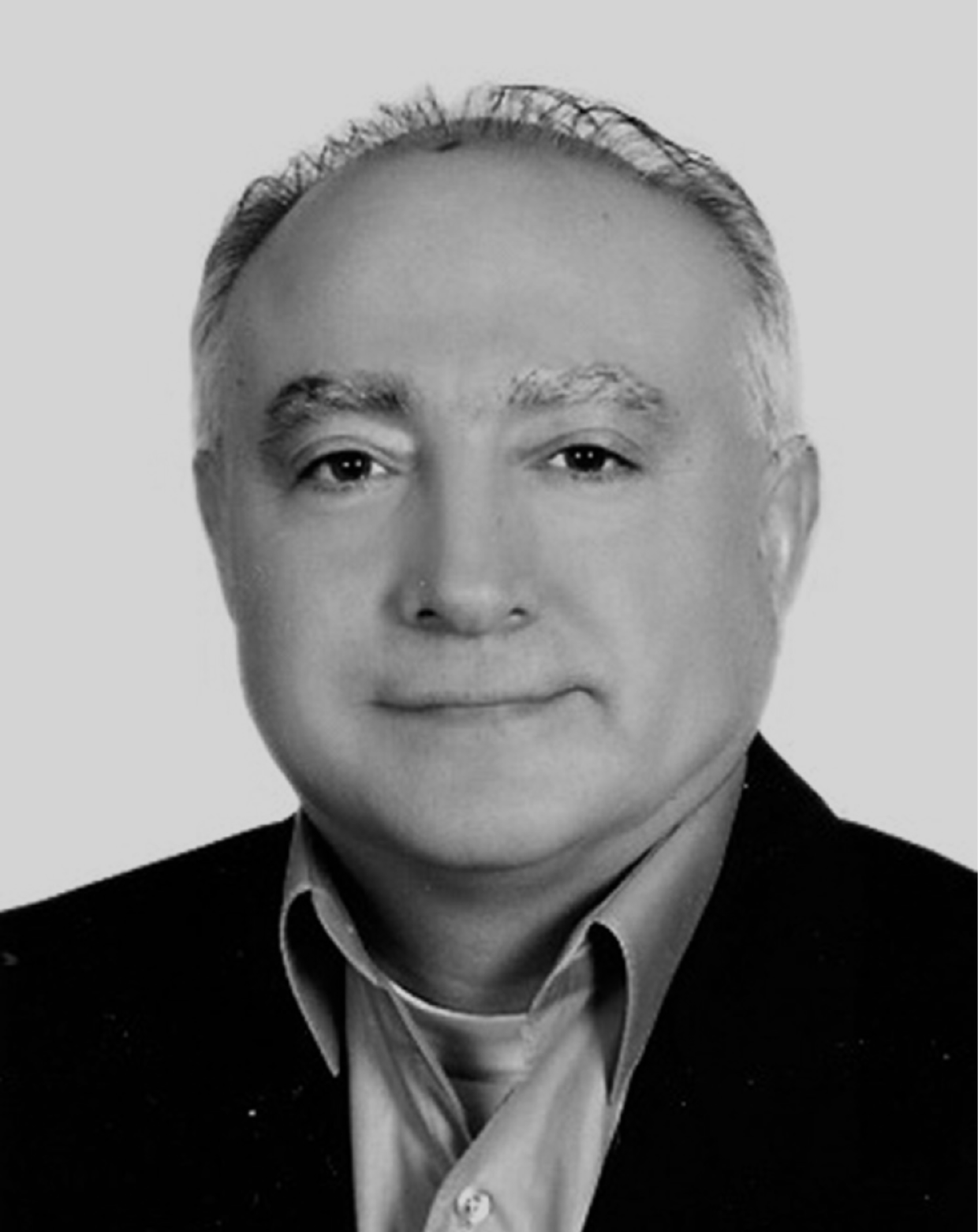}}]
	{Ali Movaghar}(Life Senior Member, IEEE) received the M.S. and Ph.D. degrees in computer, information, and control engineering from the University of Michigan, Ann Arbor, MI, USA, in 1979 and 1985, respectively. He is an Emeritus Professor in the Department of Computer Engineering at Sharif University of Technology, Tehran, Iran, where he has been a faculty member since 1993. Currently, he is a Visiting Professor at the University of Michigan, Ann Arbor, and has previously held visiting positions at the Institut National de Recherche en Informatique et en Automatique (INRIA), Paris, France, in 1984, and in the Department of Electrical Engineering and Computer Science at the University of California, Irvine, in 2011. He also worked with AT\&T Information Systems in Naperville, Illinois in 1985-1986, and taught with the University of Michigan, Ann Arbor, in 1987-1989. He is a senior member of the ACM. His research interests include performance and dependability modeling, formal verification of wireless networks, distributed real-time systems, and Internet of Things.
\end{IEEEbiography}

\thispagestyle{empty}
\begin{textblock}{19.1}(1,1.4)
	\vspace{-5mm}
	\noindent \scriptsize \hspace{4mm}
	3130 \hfill IEEE TRANSACTIONS ON NETWORK SCIENCE AND ENGINEERING, VOL. 12, NO. 4, JULY/AUGUST 2025
\end{textblock}

\end{document}